\documentclass[12pt]{article}
\usepackage[english,activeacute]{babel}
\usepackage{natbib}
\usepackage{comment}
\usepackage{float}
\usepackage[hidelinks]{hyperref}
\usepackage{mathrsfs}
\usepackage{enumitem}
\usepackage[font={small,it}]{caption}
\usepackage{amsmath,amsfonts,amsthm,amssymb}
\usepackage{bm,rotating,multirow,dsfont,graphicx}
\usepackage[usenames, dvipsnames]{color}
\usepackage{url}
\usepackage{multicol}
\usepackage{multirow}
\usepackage[T1]{fontenc}
\usepackage{flafter}
\usepackage{appendix}
\usepackage{subfigure}
\usepackage{xcolor}
\makeatletter
\def\hlinewd#1{%
	\noalign{\ifnum0=`}\fi\hrule \@height #1 %
	\futurelet\reserved@a\@xhline}
\makeatother
\newcommand\simiid{\mathrel{\overset{\makebox[0pt]{\mbox{\normalfont\tiny\sffamily iid}}}{\sim}}}
\newcommand\simind{\mathrel{\overset{\makebox[0pt]{\mbox{\normalfont\tiny\sffamily ind}}}{\sim}}}

\newcommand{\ind}[1]{\mathbf{1}\left\{ #1 \right\}}
\newcommand\floor[1]{\lfloor#1\rfloor}
\newcommand{\expec}[1]{\textsf{E}\left(#1\right)}

\newcommand{\bs}{\boldsymbol}
\def\yij{y_{i,j}}

\def\Xv{\boldsymbol{X}}

\def\xv{\boldsymbol{x}}
\def\yv{\boldsymbol{y}}

\def\muv{\boldsymbol{\mu}}
\def\Ima{\mathbf{I}}
\def\Am{\mathbf{A}}

\def\Xm{\mathbf{X}}
\newcommand{\Sm}{\mathbf{S}}

\def\Sig{\mbox{\boldmath$\Sigma$}}
\def\SIG{\mbox{\boldmath$\Sigma$}}
\def\Lam{\mbox{\boldmath$\Lambda$}}
\def\be{\beta}
\def\bev{\boldsymbol{\beta}}

\def\muv{\boldsymbol{\mu}}

\def\omev{\boldsymbol{\omega}}
\def\si{\sigma}
\def\sicv{\boldsymbol{\sigma^2}}

\def\be{\beta}

\def\al{\alpha}
\def\ep{\epsilon}

\def\ga{\gamma}
\def\trans{\textsf{T}}
\def\zerov{\boldsymbol{0}}

\def\DIC{\text{DIC}}
\def\WAIC{\text{WAIC}}
\def\@roman#1{\romannumeral #1}

\begin{document}

\def\spacingset#1{\renewcommand{\baselinestretch}{#1}\small\normalsize}\spacingset{1}

\title{A Gentle Introduction to Bayesian Hierarchical Linear
	Regression Models}

\date{}
	
\author{
	Juan Sosa, Universidad Nacional de Colombia, Colombia\footnote{jcsosam@unal.edu.co} \\
	Jeimy Aristizabal, Universidad Externado de Colombia, Colombia\footnote{jparistizabalr@gmail.com}
}	 

\maketitle

\begin{abstract} 
Considering the flexibility and applicability of Bayesian modeling, in this work we revise the main characteristics of two hierarchical models in a regression setting. We study the full probabilistic structure of the models along with the full conditional distribution for each model parameter. Under our hierarchical extensions, we allow the mean of the second stage of the model to have a linear dependency on a set of covariates. The Gibbs sampling algorithms used to obtain samples when fitting the models are fully described and derived. In addition, we consider a case study in which the plant size is characterized as a function of nitrogen soil concentration and a grouping factor (farm).
\end{abstract}


\spacingset{1.1} % DON'T change the spacing!

\newpage

\section{Introduction}\label{sec_intro}

A key characteristic of many problems is that the observed data can be used to estimate aspects of the population even though they are never observed. Often, it is quite natural to model such a problem hierarchically, with observable outcomes modeled conditionally on certain parameters, which themselves are assigned a probabilistic specification in terms of further random quantities. 

Hierarchical models can have enough parameters to fit the data well, while using a population distribution to structure some dependence into the parameters, thereby avoiding problems of over-fitting. In addition, by establishing hierarchies we are not forced to choose between complete pooling and not pooling at all as the classic analysis of variance does \citep{gelman2013bayesian}.

In this work we analyze observational continuous data arranged in groups. Specifically, we discuss hierarchical models for the comparison of group-specific parameters across groups in a regression setting. Our hierarchical approach is conceptually a straightforward generalization of a standard Normal model. Emulating \cite{hoff2009first}, we use an ordinary regression model to describe within-group heterogeneity of observations, and also, describe between-group heterogeneity using a sampling model for the group-specific regression parameters. Then, we take a step further and develop another hierarchical model adding a specific layer for carrying out clustering tasks. 

Even though our modeling approach is quite conventional under the Bayesian paradigm, our main contribution mainly relies in how models are structured and developed. Thus, we provide for each model all the details regarding model and (hierarchical) prior specification, careful hyperparameter selection under little external information, and simulation-based algorithms for computation. Finally, we also consider several metrics to quantify the performance aspects of our models by means of both goodness-of-fit and in-sample model checking metrics. In every instance along the way, we go over several buildying blocks of the literature about hierarchical modeling techniques and clustering tasks.

This paper is structured as follows: Section \ref{sec_normal_linar_regression_model} revises all the details related to the Normal linear regression model. Sections \ref{sec_hierarchical_normal_model} and \ref{sec_clustering_model} provide a full development of the hierarchical extensions. Section \ref{sec_extensions} discusses in depth other modeling approaches. Section \ref{sec_model_check} presents several specifics about model checking through tests statistics, and also, model section through information criteria. Section \ref{sec_application} makes a complete analyzes of a case study. Finally, Section \ref{sec_discussion} discusses our findings and future developments.

\newpage

\section{Normal Linear Regression Model}
\label{sec_normal_linar_regression_model}

Here, we show some relevant aspects about linear regression modeling in a Bayesian setting, which is a powerful data analysis tool quite useful for carrying out many inferential tasks such as data characterization and prediction. Roughly speaking, our goal is to find a model for predicting the dependent variable (response) given one or more independent (predictor) variables.

\subsection{Model Specification}

First, we consider a simple scenario in which we want to characterize the sampling distribution of a random variable $y$ through a set of explanatory variables $\xv=[x_1,\ldots,x_p]^{\textsf{T}}$. 
Thus, we look upon a Normal linear regression model of the form
\begin{gather*}
    \yij = \xv_{i,  j}^{\textsf{T}}\bev + \ep_{i,j}\,, \quad \ep_{i,j}\mid\si^2\simiid \textsf{N}(0,\si^2)\,, \quad
    i=1,\ldots,n\,, \quad 
    j=1,\ldots,m\,,
\end{gather*}
where $\yij$, $\xv_{i,j}$, and $\ep_{i,j}$ are the response variable, the covariates, and the random error, respectively, corresponding to the $i$-th observation from the $j$-th group, and $\bev = [\be_1,\ldots,\be_p]^{\textsf{T}}$ are the regression parameters of the model. Note that the previous model can be re-expressed as
\begin{gather*}
    \yv\mid\Xm,\bev,\si^2\sim \textsf{N}_{nm}(\Xm\bev,\si^2\Ima)
\end{gather*}
where $\yv$ is the response vector given by $\yv=[\yv_1^{\textsf{T}},\ldots,\yv_m^{\textsf{T}}]^{\textsf{T}}$, with $\yv_j=[y_{j,1}\ldots,y_{j,n}]^{\textsf{T}}$, and $\Xm$ is the design matrix arranged in a similar fashion.

In order to perform full Bayesian analysis using the likelihood given above, we consider a semi-conjugate prior distribution for $\bev$ and $\si^2$ of the form
$$
\bev\sim \textsf{N}(\bev_0, \Sig_0)
\quad\text{and}\quad
\si^2\sim \textsf{IG}(\nu_0/2, \nu_0\si_0^2/2)
$$
as in most Normal sampling problems. 

\subsection{Prior Elicitation}

In the absence of convincing external information to the data, it is customary using a defuse prior distribution in order to be as minimally informative as possible. In the same spirit of the unit information prior as in \cite{kass1995reference}, we let $g = nm$ and $\nu_0 = 1$, and set $\bev_0 = \hat{\bs{\beta}}_{\text{OLS}}$, $\SIG_0 = g\,\si_{0}^2(\Xm^{\trans}\Xm)^{-1}$, with $\sigma^2_0 = \hat\si^2_{\text{OLS}}$, where $\hat\phi_{\text{OLS}}$ stands for the ordinary least squares (OLS) estimate of $\phi$. This choice of $g$ makes the ratio $\frac{g}{g+1}$ very close to 1, and therefore, $\bev$ is practically centered around $\bs{0}$; similarly, the prior distribution of $\si^2$ is weakly centered around $\hat{\si}^2_{\text{OLS}}$ since $\nu_0=1$. Such a prior distribution cannot be strictly considered a real prior distribution, as it requires knowledge of $\yv$ to be constructed. However, it only uses a small amount of the information given in $\yv$, and can be loosely thought of as the prior distribution of a researcher with
unbiased but weak prior information. 

\subsection{Posterior Inference}

The posterior distribution can be explored using Markov chain Monte Carlo (MCMC) methods \citep{gamerman2006markov} such as the Gibbs sampling. Implementing such an algorithm under the previous model is quite simple since the full conditional distributions are
$$
\bev\mid\yv,\Xm,\si^2 \sim \textsf{N}_p((\Sig_0^{-1}+\si^{-2}\Xm^{\textsf{T}}\Xm)^{-1}(\Sig_0^{-1}\bev_0 + \si^{-2}\Xm^{\textsf{T}}\yv) , (\Sig_0^{-1}+\si^{-2}\Xm^{\textsf{T}}\Xm)^{-1})
$$
and
$$
\si^2\mid\yv,\Xm,\bev \sim \textsf{IG}((\nu_0+nm)/2, (\nu_0\si^2_0 + (\yv-\Xm\bev)^{\textsf{T}}(\yv-\Xm\bev))/2)\,.
$$
See for example \citet{christensen2011bayesian} for details about this result.

Under the previous setting, we need to carefully choose values for the set of model hyperparameters, namely, $\bev_0$, $\Sig_0$, $\nu_0$, and $\si^2_0$. Often, the analysis must be done in the absence of prior information, so we should use a prior distribution as minimally informative as possible. The so-called $g$-priors (see for example \citealp{albert2009bayesian} for a brief discussion) offer this possibility and the desirable feature of invariance to changes in the scale of the regressors. A popular alternative in this direction consists in letting $\bev_0 = \boldsymbol{0}$ and $\Sig_0=g\si^2(\Xm^{\textsf{T}}\Xm)^{-1}$ for some positive value $g$ that reflects the amount of information in the data relative to the prior distribution (choosing a large value of $g$ naturally induces a diffuse prior). It can be shown that under the $g$-prior specification, $p(\si^2\mid\yv,\Xm)$ is an Inverse Gamma distribution, which means that we can use direct sampling as follows (see \citealp{hoff2009first} for details):
\begin{enumerate}
  \item Sample
  $\si^2\sim \textsf{IG}\left((\nu_0+nm)/2, (\nu_0\si^2_0 + \yv^{\textsf{T}}(\Ima-\tfrac{g}{g+1}\Xm(\Xm^{\textsf{T}}\Xm)^{-1}\Xm^{\textsf{T}})\yv)/2\right)$.
  \item Sample $\bev\sim \textsf{N}_p\left(\frac{g}{g+1}(\Xm^{\textsf{T}}\Xm)^{-1}\Xm^{\textsf{T}}\yv, \frac{g}{g+1}\sigma^2(\Xm^{\textsf{T}}\Xm)^{-1}\right)$.
\end{enumerate}

\section{Hierarchical Normal Linear Regression Model} \label{sec_hierarchical_normal_model}

We present the treatment of a hierarchical model, in which the observed data is assumed to be normally distributed with both group-specific fixed effects (and therefore subject-specif means) and group-specific variances. The model specification provided below is quite convenient because in addition to a global assessment of the mean relationship between the covariates and the response variable (as allowed by a standard linear regression model), it gives the means to carry out specific inferences within each group as well as comparisons among groups. 

\subsection{Model Specification}

\begin{figure}[!b]
\centering
\includegraphics[scale=.7]{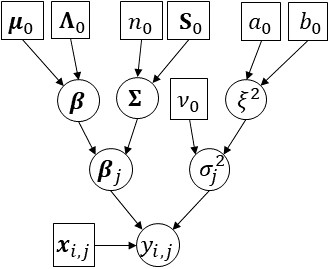}
\caption{DAG representation of the hierarchical Normal linear regression model.}
\label{fig_diagram}
\end{figure}

We consider $m$ independent groups, each one of them with $n$ independent normally distributed data points (i.e., a balanced experiment), $\yij$, each of which with subject-specific mean $\mu_{i,j}=\xv_{i,j}^{\textsf{T}}\bev_j$, with $\bev_j = (\beta_{1,j},\ldots,\beta_{p,j})$, and group-specific variance $\si_j^2$; i.e.,
$$
\yij\mid\xv_{i,j},\bev_j,\si^2_j\simind \textsf{N}\left(\xv_{i,j}^{\textsf{T}}\bev_j\,,\si^2_j\right)\,,
\quad i=1,\ldots,n\,,\quad j=1,\ldots,m\,.
$$
In addition, we propose a hierarchical prior distribution with the following stages:
$$
\bev_j\mid\bev, \SIG \simiid \textsf{N}_p (\bev, \SIG)
\quad\text{and}\quad 
\si_j^2\mid\xi^2 \simiid \textsf{IG}(\nu_0/2,\nu_0\xi^2/2)\,,
$$
with
$$
\bev   \sim \textsf{N}_p(\bs{\mu}_0,\mathbf{\Lambda}_0)\,,\quad
\SIG   \sim \textsf{IW}(n_0, \Sm^{-1}_0)\,,\quad
\xi^2 \sim \textsf{G}(a_0,b_0)\,,
$$
where $\bs{\zeta}=(\bev_{1},\ldots,\bev_{m})$, $\sicv=(\si^2_1,\ldots,\si^2_m)$, $\bev$, $\Sig$, and $\xi^2$ are the unknown model parameters, and $\bs{\mu}_0$, $\mathbf{\Lambda}_0$, $n_0$, $\Sm_0$, $\nu_0$, $a_0$, and $b_0$ are hyperparameters carefully selected according to external information. Figure \ref{fig_diagram}) provides a directed acyclic graph (DAG) representation of the model. As a final remark, note that fitting this hierarchical model is not equivalent to fitting regular regression models to each group independently since the information shared across groups (shrinkage effect) would be lost. See \citet{hoff2009first} for more details in this regard.

\subsection{Prior Elicitation}

Following the same unit-information-prior-inspired approach considered to select the hyperparameters of the Normal linear regression model, we let again $g = nm$ and $\nu_0 = 1$, and set $\muv_0 = \hat{\bs{\beta}}_{\text{OLS}}$, $\mathbf{\Lambda}_0 = g\,\si_{0}^2(\Xm^{\trans}\Xm)^{-1}$, with $\sigma^2_0 = \hat\si^2_{\text{OLS}}$ (see Section \ref{sec_normal_linar_regression_model} for details). In addition, aiming to establishing a diffuse and reasonable centered prior for $\mathbf{\Sigma}$, we let $n_0=p+2$ and $\mathbf{S}_0=\mathbf{\Lambda}_0$ because such a specification produces a mean vague concentration of $\bev$ around $\muv_0$ since $\textsf{E}(\SIG)=\mathbf{\Lambda}_0$ a priori. Finally, we let $a_0=1$ and $b_0=1/\sigma^2_0$ because this choice leads to a diffuse prior for $\xi^2$ such that $\textsf{E}(\xi^2)=\sigma_0$ with $\textsf{CV}(\xi^2)=1$, which clearly emulates the prior elicitation in a regular regression setting for which $\textsf{E}(\sigma^2)=\sigma^2_0$. 

\subsection{Posterior Inference}\label{sec_posterior_inference}

Joint posterior inference for the model parameters can be achieved by constricting a Gibbs sampling algorithm \citep{gamerman2006markov}, which requires iteratively sampling each parameter from its full conditional distribution.

%As before, let $\yv_j=[y_{j,1},\ldots,y_{j,n}]^{\textsf{T}}$ be the measurements associated with group $j$, for $j=1,\ldots,m$, $\yv=[\yv_1^{\textsf{T}},\ldots,\yv_m^{\textsf{T}}]^{\textsf{T}}$ be all the observations available in the experiment, $\Xm$ be the design matrix arranged in a similar fashion as $\yv$, and 
Let $\mathbf{\Theta}=(\bs{\zeta},\sicv,\bev,\Sig,\xi^2)$ be the full set of parameters in the model. The posterior distribution of $\mathbf{\Theta}$ is
$$
p(\mathbf{\Theta}\mid\yv,\Xm) \propto p (\yv\mid\Xm,\bs{\zeta},\sicv) \,p(\bs{\zeta}\mid\bev,\Sig)\,p(\bev)\,p(\Sig)\,p (\sicv\mid\xi^2)\,p(\xi^2)\,,
$$ 
which leads to
\begin{align*}
p(\mathbf{\Theta}\mid\yv,\Xm) &\propto
\prod_{j=1}^m \prod_{i=1}^n \si_j^{-1/2}\exp{\left\{-\tfrac{1}{2\si_j^2}\left(\yij-\xv_{i,j}^{\textsf{T}}\bev_j\right)^2\right\}}\\
&\hspace{0.5cm}\times \prod_{j=1}^m |\Sig|^{-1/2}\exp{\left\{-\tfrac12(\bev_j-\bev)^{\textsf{T}}\Sig^{-1}(\bev_j-\bev)\right\}} \\ 
&\hspace{0.5cm}\times
\exp{\left\{-\tfrac12(\bev-\muv_0)^{\textsf{T}}\Lam_0^{-1}(\bev-\muv_0)\right\}} 
\times |\mathbf{\Sigma}|^{-(n_0+p+1)/2}\,\exp{\left\{-\tfrac12\text{tr}(\Sm_0\mathbf{\Sigma}^{-1}) \right\}} \\
&\hspace{0.5cm}\times \prod_{j=1}^m(\xi^2)^{\nu_0/2}(\si_j^2)^{-(\nu_0/2+1)}\exp{\left\{-\tfrac{\nu_0\xi^2/2}{\si_j^2}\right\}}
\times (\xi^2)^{a_0-1}\exp{\{-b_0\xi^2\}}\,.
\end{align*}
Although this is an abuse of standard mathematical notation, the full conditional distribution (fcd) of parameter $\phi$ given the rest of the parameters, the design matrix $\Xm$, and the data $\yv$ is denoted by $p(\phi\mid\text{rest})$. We derived these distributions looking at the dependencies in the full posterior distribution. Thus, we have that:
\begin{itemize}

\item The fcd of $\bev_j$, for $j=1,\ldots,m$, is
\begin{align*}
    \bev_j\mid\text{rest} \sim 
    \textsf{N}_{p}\left( \left(\SIG^{-1} + \sigma_j^{-2}\Xm_j^{\trans}\Xm_j \right)^{-1} \left( \SIG^{-1}\bev + \sigma_j^{-2}\Xm_j^{\trans}\yv_j \right) , \left(\SIG^{-1} + \sigma_j^{-2}\Xm_j^{\trans}\Xm_j \right)^{-1} \right)
\end{align*}
where $\Xm_j= [\xv_{1,j},\ldots,\xv_{n,j}]^{\textsf{T}}$.

\item The fcd for $\bev$ is
\begin{equation*}
    \bev\mid\text{rest} \sim \textsf{N}_p\left( \left(\Lam_0^{-1} + m\SIG^{-1} \right)^{-1} \left( \Lam_0^{-1}\muv_0 + \SIG^{-1}\textstyle\sum_{j=1}^m \bev_j \right) , \left(\Lam_0^{-1} + m\SIG^{-1} \right)^{-1} \right)\,.
\end{equation*}

\item The fcd of $\Sig$ is
\begin{equation*}
    \Sig\mid\text{rest} \sim \textsf{IW}\left(n_0 + m, \left( \Sm_0 + \textstyle\sum_{j=1}^m (\bev_j - \bev)(\bev_j - \bev)^{\trans} \right)^{-1} \right)\,.
\end{equation*}

\item The fcd of $\si_j^2$, for $j=1,\ldots,m$, is
\begin{equation*}
    \si_j^2\mid\text{rest} \sim \textsf{IG}\left((\nu_0+n)/2 , \left(\nu_0\xi^2 + \textstyle\sum_{i=1}^n (\yij-\xv_{i,j}^{\textsf{T}}\bev_j)^2\right)/2 \right)\,.
\end{equation*}

\item The fcd of $\xi^2$ is
$$
\xi^2\mid\text{rest} \sim \textsf{G}\left( a_0 + m\nu_0/2, b_0 + \tfrac{\nu_0}2\textstyle\sum_{j=1}^m \si_j^{-2} \right) \,.
$$
\end{itemize}

Let $\phi^{(b)}$ denote the state of parameter $\phi$ in the $b$-th iteration of the Gibbs sampling algorithm, for $b=1,\ldots,B$. Then, such an algorithm in this case is as follows:
  \begin{enumerate}
    \item Choose an initial configuration for each parameter in the model, say $\bev_1^{(0)},\ldots,\bev_m^{(0)}$, $\bev^{(0)}$, $\Sig^{(0)}$, $(\si_1^2)^{(0)},\ldots,(\si_m^2)^{(0)}$, and $(\xi^2)^{(0)}$.
    \item Update $\bev_1^{(b-1)},\ldots,\bev_m^{(b-1)}$, $\bev^{(b-1)}$, $\Sig^{(b-1)}$, $(\si_1^2)^{(b-1)},\ldots,(\si_m^2)^{(b-1)}$, and $(\xi^2)^{(b-1)}$ until convergence:
    \begin{enumerate}
        \item Sample $\bev_j^{(b)}$ from the fcd
        $p(\bev_j\mid(\si_j^2)^{(b-1)},\bev^{(b-1)},\Sig^{(b-1)},\yv_j,\Xm_j)$, for $j=1,\ldots,m$. 
        \item Sample $\bev^{(b)}$ from the fcd
        $p(\bev\mid\{\bev_j^{(b)}\},\Sig^{(b-1)}).$
        \item Sample $\Sig^{(b)}$ from the fcd
        $p(\Sig\mid\{\bev_j^{(b)}\},\bev^{(b)}).$
        \item Sample $(\si_j^2)^{(b)}$ from the fcd
        $p(\si_j^2\mid\bev_j^{(b)},(\xi^2)^{(b-1)},\yv_j,\Xm_j)$, for $j=1,\ldots,m$.
      \item Sample $(\xi^2)^{(b)}$ from the fcd
      $p(\xi^2\mid\{(\si^2_j)^{(b)}\}).$
    \end{enumerate}
    \item Cycle until achieve convergence.
  \end{enumerate}

\section{Clustering Hierarchical Normal Linear Regression Model} \label{sec_clustering_model}

Here we extend the hierarchical approach provided in the previous section by adding more structure to the model in order to perform clustering task at a group level (clusters whose elements are groups). Specifically, we consider a mixture model instead of a regular Normal distribution in the likelihood, and then, introduce a new set of parameters known as cluster assignments to ``break'' the mixture and then be able to identify those groups belonging to the same cluster. 

\subsection{Model Specification}

A natural way to extend the hierarchical model consists in relaxing the normality assumption about the response variable by replacing it with a finite mixture of Normal components in such a way that
$$
y_{i,j}\mid \xv_{i,j}, \{\bev_k\} \{\sigma^2_k\} \simind \sum_{k=1}^K \omega_k \textsf{N}\left(y_{i,j}\mid\xv_{i,j}^{\textsf{T}}\bev_k, \sigma^2_k\right)\,,
\quad i=1,\ldots,n\,,\quad j=1,\ldots,m\,,
$$
where $K$ is a positive fixed integer that represents the number of clusters in which groups can be classified, $\bev_1,\ldots,\bev_K$ and $\si^2_1,\ldots,\si^2_K$ are the cluster-specific regression parameters and cluster-specific variances of the mixture components, and $\omega_1,\ldots,\omega_K$ are mixture probabilities such that $0<\omega_k<1$ and $\sum_{k=1}^K\omega_k = 1$. Note that under this formulation we recover the Normal linear regression model by setting $K=1$.

According to the previous mixture, the probability that the group $j$ is part of the cluster $k$ is $\omega_k$, i.e., $\textsf{Pr}(\ga_j = k \mid \omega_k) = \omega_k$, for $j=1,\ldots,m$ and $k=1,\ldots,K$, where $\ga_j$ is a categorical variable (known as either cluster assignment or cluster indicator) that takes integer values in $1,\ldots,K$ with probabilities $\omega_1,\ldots,\omega_K$, respectively. Thus, we can use the cluster assignments $\ga_1,\ldots,\ga_m$ to ``break'' the mixture and write the model as
$$
y_{i,j}\mid \xv_{i,j}, \ga_j, \bev_{\ga_j},\sigma^2_{\ga_j} \simind \textsf{N}\left(y_{i,j}\mid\xv_{i,j}^{\textsf{T}}\bev_{\ga_j}, \sigma^2_{\ga_j}\right)
\quad i=1,\ldots,n\,,\quad j=1,\ldots,m\,.
$$
In addition, a parsimonious way to formulate a hierarchical prior distribution can be achieved by letting
$$
\bs{\ga}\mid\omev \sim \textsf{Cat}(\omev)\,,\quad
\bev_k\mid\bev, \SIG \simiid \textsf{N}_p (\bev, \SIG)
\quad\text{and}\quad 
\si_k^2\mid\xi^2 \simiid \textsf{IG}(\nu_0/2,\nu_0\xi^2/2)\,,
$$
with
$$
\omev\sim \textsf{Dir}(\bs{\alpha}_0)\,,\quad
\bev   \sim \textsf{N}_p(\bs{\mu}_0,\mathbf{\Lambda}_0)\,,\quad
\SIG   \sim \textsf{IW}(n_0, \Sm^{-1}_0)\,,\quad
\xi^2 \sim \textsf{G}(a_0,b_0)\,,
$$
where $\bs{\ga} = (\ga_1,\ldots,\ga_m)$, $\bs{\omega}=(\omega_1,\ldots,\omega_K)$, $\bs{\zeta}=(\bev_{1},\ldots,\bev_{K})$, $\sicv=(\si^2_1,\ldots,\si^2_K)$, $\bev$, $\Sig$, and $\xi^2$ are the unknown model parameters, and $\bs{\alpha}_0$, $\bs{\mu}_0$, $\mathbf{\Lambda}_0$, $n_0$, $\Sm_0$, $\nu_0$, $a_0$, and $b_0$ are hyperparameters carefully selected according to external information. Finally, note that the DAG representation of the model is very similar to that displayed in Figure \ref{fig_diagram}, but including a few extra nodes corresponding to the clustering process.

\subsection{Prior Elicitation}

Under the same approach as before, we let once again $g = nm$ and $\nu_0 = 1$, and set $\muv_0 = \hat{\bs{\beta}}_{\text{OLS}}$, $\mathbf{\Lambda}_0 = g\,\si_{0}^2(\Xm^{\trans}\Xm)^{-1}$, with $\sigma^2_0 = \hat\si^2_{\text{OLS}}$, $n_0=p+2$, $\mathbf{S}_0=\mathbf{\Lambda}_0$, $a_0=1$, and $b_0=1/\sigma^2_0$. Thus, the only hyperparameter that remains unspecified is $\bs{\alpha^0}$. To do so in a sensible way, we let $\alpha^0=\left(\tfrac{1}{K},\ldots,\tfrac{1}{K}\right)$, which has a
direct connection with a Chinese restaurant process prior \citep{ishwaran2000markov} and places a diffuse prior distribution for the number of occupied clusters in the data.

\subsection{Posterior Inference}

Once again we recur to MCMC methods as in Section \ref{sec_hierarchical_normal_model} to explore the posterior distribution of the model parameters $\mathbf{\Theta}=(\bs{\ga},\bs{\omega},\bs{\zeta},\sicv,\bev,\Sig,\xi^2)$.
The posterior distribution of $\mathbf{\Theta}$ is such that
$$
p(\mathbf{\Theta}\mid\yv,\Xm) \propto p (\yv\mid\Xm,\bs{\ga},\bs{\zeta},\sicv) \,p(\bs{\ga}\mid\bs{\omega})\,p(\bs{\omega})\,p(\bs{\zeta}\mid\bev,\Sig)\,p(\bev)\,p(\Sig)\,p (\sicv\mid\xi^2)\,p(\xi^2)\,,
$$ 
which leads to
\begin{align*}
p(\mathbf{\Theta}\mid\yv,\Xm) &\propto
\prod_{j=1}^m \prod_{i=1}^n \si_j^{-1/2}\exp{\left\{-\tfrac{1}{2\si_{\ga_j}^2}\left(\yij-\xv_{i,j}^{\textsf{T}}\bev_{\ga_j}\right)^2\right\}}\times \prod_{j=1}^m\prod_{k=1}^K \omega_k^{[\ga_j=k]} \times \prod_{k=1}^K \omega_k^{\alpha_{0k}} \\
&\hspace{0.5cm}\times \prod_{k=1}^K |\Sig|^{-1/2}\exp{\left\{-\tfrac12(\bev_k-\bev)^{\textsf{T}}\Sig^{-1}(\bev_k-\bev)\right\}} \\ 
&\hspace{0.5cm}\times
\exp{\left\{-\tfrac12(\bev-\muv_0)^{\textsf{T}}\Lam_0^{-1}(\bev-\muv_0)\right\}} 
\times |\mathbf{\Sigma}|^{-(n_0+p+1)/2}\,\exp{\left\{-\tfrac12\text{tr}(\Sm_0\mathbf{\Sigma}^{-1}) \right\}} \\
&\hspace{0.5cm}\times \prod_{k=1}^K(\xi^2)^{\nu_0/2}(\si_k^2)^{-(\nu_0/2+1)}\exp{\left\{-\tfrac{\nu_0\xi^2/2}{\si_k^2}\right\}}
\times (\xi^2)^{a_0-1}\exp{\{-b_0\xi^2\}}\,,
\end{align*}
where $[x=i]$ is the Iverson bracket. Such a posterior distribution is quite reminiscent of the one that we derived for the hierarchical model in Section \ref{sec_hierarchical_normal_model}, but this time it includes a portion related with the clustering process, and also, group-specific parameters cycle over $K$ terms instead of $m$.

Once again, we derive the fcd's from the posterior distribution, obtaining that:
\begin{itemize}

\item The fcs of $\ga_j$, for $j=1,\ldots,m$, is a Categorical distribution such that
$$\textsf{Pr}(\ga_j = k\mid\text{rest})\propto \omega_k\prod_{i=1}^n \textsf{N}(y_{i,j}\mid \xv_{i,j}^{\textsf{T}}\bev_k,\si^2_k)\,,\quad\text{for $k =1,\ldots,K$\,.}$$

\item The fcs of $\bs{\omega}$ is
$$
\bs{\omega}\mid\text{rest}\sim \textsf{Dir}(\al_{01}+n_1,\ldots,\al_{0K}+n_K)
$$
where $n_k=\#\{j:\ga_j=k\}$ is the number of elements in cluster $k$, for $k= 1,\ldots,K$.

\item The fcd of $\bev_k$, for $k=1,\ldots,K$, is
\begin{align*}
    \bev_k\mid\text{rest} \sim 
    \textsf{N}_{p}\left( \left(\SIG^{-1} + \sigma_k^{-2}\Xm_{(k)}^{\trans}\Xm_{(k)} \right)^{-1} \left( \SIG^{-1}\bev + \sigma_k^{-2}\Xm_{(k)}^{\trans}\yv_{(k)} \right) , \left(\SIG^{-1} + \sigma_k^{-2}\Xm_{(k)}^{\trans}\Xm_{(k)} \right)^{-1} \right)
\end{align*}
where $\Xm_{(k)}= [\Xm_j^{\textsf{T}}:\ga_j=k]^{\textsf{T}}$ and $\yv_{(k)}= [\yv_j^{\textsf{T}}:\ga_j=k]^{\textsf{T}}$. Note that if cluster $k$ is empty, then the fcd of $\bev_k$ is just $\bev_k\mid\text{rest} \sim \textsf{N}_{p}(\bev,\SIG)$. 

\item The fcd for $\bev$ is
\begin{equation*}
    \bev\mid\text{rest} \sim \textsf{N}_p\left( \left(\Lam_0^{-1} + K^*\SIG^{-1} \right)^{-1} \left( \Lam_0^{-1}\muv_0 + \SIG^{-1}\textstyle\sum_{k:n_k>0}^{K} \bev_k \right) , \left(\Lam_0^{-1} + K^*\SIG^{-1} \right)^{-1} \right)
\end{equation*}
where $K^*$ is the number of non-empty clusters.

\item The fcd of $\Sig$ is
\begin{equation*}
    \Sig\mid\text{rest} \sim \textsf{IW}\left(n_0 + K^*, \left( \Sm_0 + \textstyle\sum_{k:n_k>0}^K (\bev_k - \bev)(\bev_k - \bev)^{\trans} \right)^{-1} \right)\,.
\end{equation*}

\item The fcd of $\si_k^2$, for $k=1,\ldots,K$, is
\begin{equation*}
    \si_k^2\mid\text{rest} \sim \textsf{IG}\left((\nu_0+n_k)/2 , \left(\nu_0\xi^2 + \textstyle\sum_{j:\ga_j=k}^m\textstyle\sum_{i=1}^n \left(\yij-\xv_{i,j}^{\textsf{T}}\bev_{\ga_j}\right)^2\right)/2 \right)\,.
\end{equation*}
Again, note that if cluster $k$ is empty, then the fcd of $\si^2_k$ is just $\si^2_k\mid\text{rest} \sim \textsf{IG}(\nu_0/2,\nu_0\xi^2/2)$.

\item The fcd of $\xi^2$ is
$$
\xi^2\mid\text{rest} \sim \textsf{G}\left( a_0 + K^*\nu_0/2, b_0 + \tfrac{\nu_0}2\textstyle\sum_{k:n_k>0}^K \si_k^{-2} \right) \,.
$$
\end{itemize}

Thus, the Gibbs sampling algorithm in this case is as follows:
  \begin{enumerate}
    \item For a given value of $K$, choose an initial configuration for each parameter in the model, say $\ga_1^{(0)},\ldots,\ga_m^{(0)}$, $\bs{\omega}^{(0)}$, $\bev_1^{(0)},\ldots,\bev_K^{(0)}$, $\bev^{(0)}$, $\Sig^{(0)}$, $(\si_1^2)^{(0)},\ldots,(\si_K^2)^{(0)}$, and $(\xi^2)^{(0)}$.
    \item Update $\ga_1^{(b-1)},\ldots,\ga_m^{(b-1)}$, $\bs{\omega}^{(b-1)}$, $\bev_1^{(b-1)},\ldots,\bev_K^{(b-1)}$, $\bev^{(b-1)}$, $\Sig^{(b-1)}$, $(\si_1^2)^{(b-1)},\ldots,(\si_K^2)^{(b-1)}$, and $(\xi^2)^{(b-1)}$ until convergence:
    \begin{enumerate}
        \item Sample $\ga_j^{(b)}$ from the fcd $p(\ga_j\mid\bs{\omega}^{(b-1)},\{\bev_k^{(b-1)}\},\{(\si_k^2)^{(b-1)}\},\Xm_j)$, for $j=1,\ldots,m$.
        \item Sample $\bs{\omega}^{(b)}$ from the fcd $p(\bs{\omega}\mid\{\ga_j^{(b)}\})$.
        \item Sample $\bev_k^{(b)}$ from the fcd
        $p(\bev_k\mid\{\ga_j^{(b)}\},(\si_k^2)^{(b-1)},\bev^{(b-1)},\Sig^{(b-1)},\{\yv_j\},\{\Xm_j\})$, for $k=1,\ldots,K$. 
        \item Sample $\bev^{(b)}$ from the fcd
        $p(\bev\mid\{\bev_k^{(b)}\},\Sig^{(b-1)}).$
        \item Sample $\Sig^{(b)}$ from the fcd
        $p(\Sig\mid\{\bev_k^{(b)}\},\bev^{(b)}).$
        \item Sample $(\si_k^2)^{(b)}$ from the fcd
        $p(\si_k^2\mid\{\ga_j^{(b)}\},\{\bev_k^{(b)}\},(\xi^2)^{(b-1)},\{\yv_j\},\{\Xm_j\})$, for $k=1,\ldots,K$.
      \item Sample $(\xi^2)^{(b)}$ from the fcd
      $p(\xi^2\mid\{(\si^2_k)^{(b)}\}).$
    \end{enumerate}
    \item Cycle until achieve convergence.
  \end{enumerate}
  
\section{Further Extensions}\label{sec_extensions}

In order to gain stochastic flexibility, yet another way of extending the model provided in Section \ref{sec_hierarchical_normal_model} consists in completely relaxing the normality assumption about the response variable by assigning a prior distribution to it. Specifically, we consider a Dirichlet process (DP) mixture model of the form
$$
y_{i,j}\mid \xv_{i,j}, \sigma^2_j, G \simind \int\textsf{N}\left(y_{i,j}\mid\xv_{i,j}^{\textsf{T}}\bev_j, \sigma^2_j\right)\textsf{d}G(\bev_j)
\quad\text{and}\quad
G\sim\textsf{DP}(\alpha,H)
$$
where $\alpha$ is a positive scalar parameter and $H$ is a base distribution function. In this case, the DP generates cumulative distribution functions on $\mathbb{R}$ (see for example \citealp{muller2015bayesian} for a formal treatment of the DP). The model given above can be also written as
$$
y_{i,j}\mid \xv_{i,j}, \bev_j,\sigma^2_j \simind \textsf{N}\left(\xv_{i,j}^{\textsf{T}}\bev_j, \sigma^2_j\right)\,,
\quad
\bev_j\mid G \sim G
\quad\text{and}\quad
G\sim\textsf{DP}(\alpha,H)\,,
$$
which makes evident why the $\mu_{i,j}=\xv_{i,j}^{\textsf{T}}\bev_j$ can be interpreted as subject-specific random effects. Such an extension is beyond the scope of this work and will be discussed in detail elsewhere.

Other straightforward parametric extensions are considering group-specific effects $\theta_1,\ldots,\theta_m$ in a way that 
$\textsf{E}(y_{i,j}\mid\xv_{i,j},\bev_{j},\theta_j) = \xv_{i,j}^{\textsf{T}}\bev_j + \theta_j$, for $i=1,\ldots,n$, $j=1,\ldots,m$,
as well as extra model hierarchies such as letting $\nu_0$ to be a integer random value ranging from 1 to a fixed large upper bound in a way that $p(\nu)\propto e^{-\kappa_0\nu}$, where $\kappa_0$ is a hyperparameter. For clustering tasks, more sophisticated extensions require the specification of nonparametric priors of the form
$\gamma_j\mid\{\omega_k\} \simiid \sum_{k=1}^\infty \omega_k\delta_k$, for $j=1,\ldots,m$,
where $\omega_k = u_k\prod_{h<k}(1-u_h)$ are weights constructed from a sequence $u_1,u_2,\ldots$, with $u_k\simind \textsf{Beta}(1-a,b+ka)$ for $0<a<1$ and $b > -a$. The joint distribution of the set of weights $\omega_1,\omega_2,\ldots$ is called a stick-breaking prior with parameters $a$ and $b$. This formulation is connected to the stick-breaking construction of the Poisson-Dirichlet process \citep{pitman1997two}. The stick-breaking representation associated with the Dirichlet process is a special case with $a = 0$.  

\begin{figure}[!b]
\centering
\subfigure[Linear Regression Model]{\includegraphics[scale=0.65]{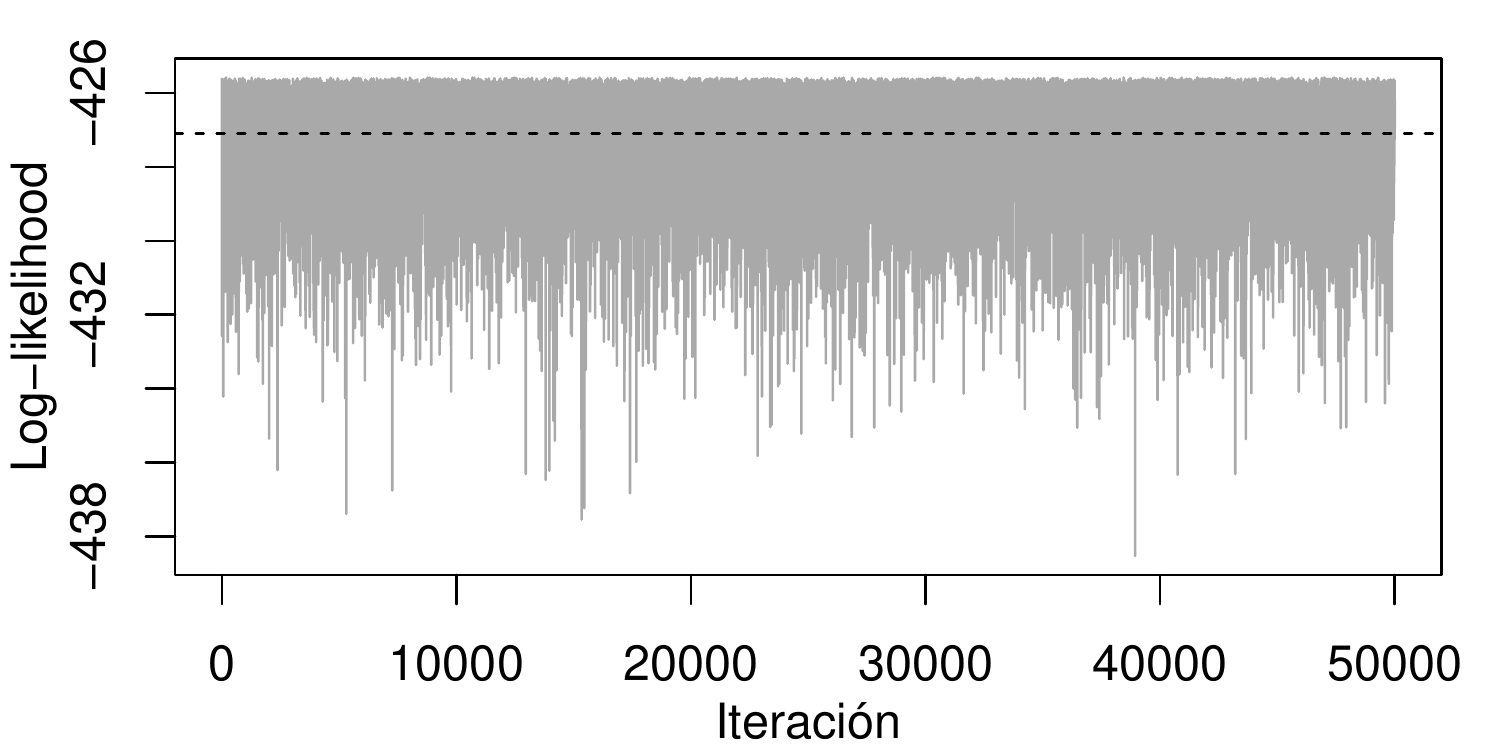}}
\subfigure[Hierarchical Linear Regression Model]{\includegraphics[scale=0.65]{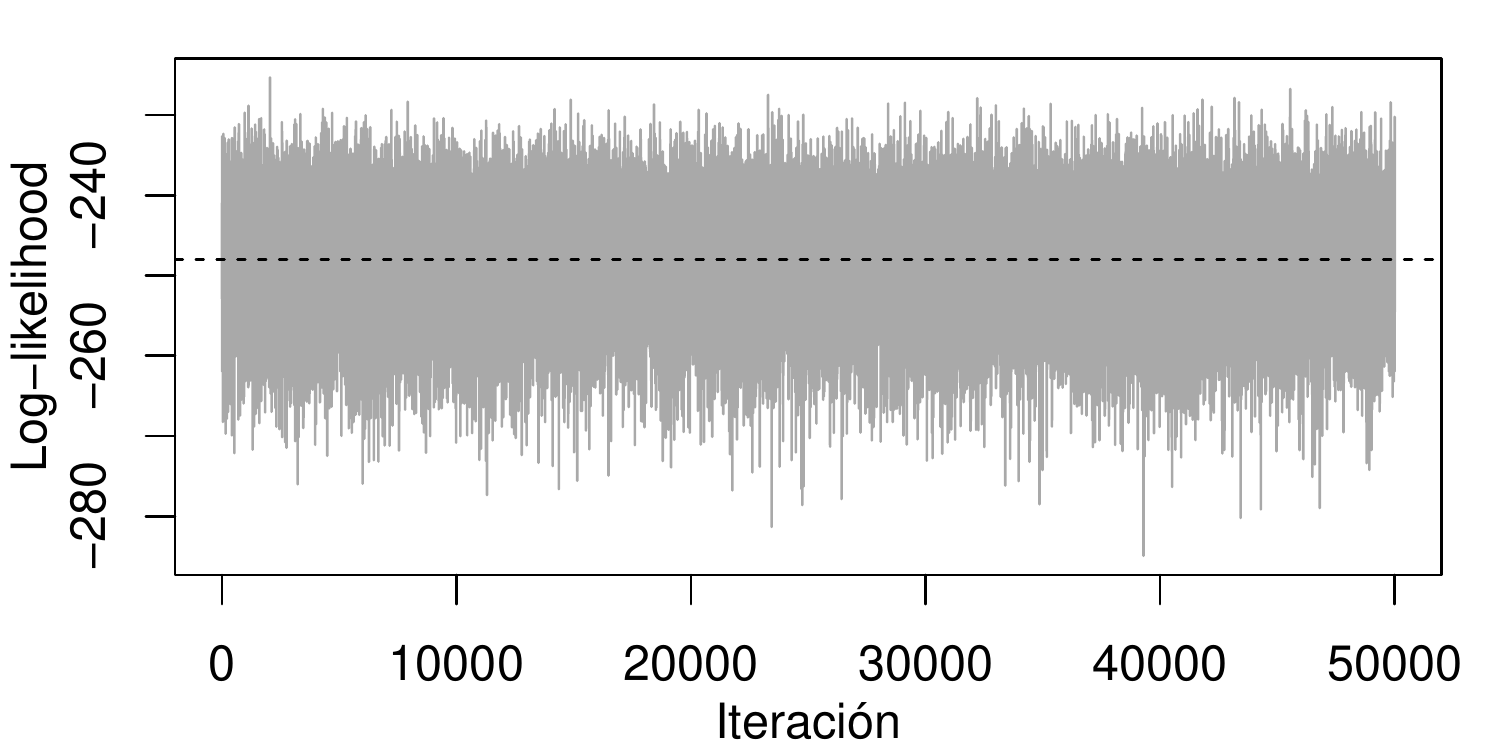}}
\subfigure[Clustering Hierarchical Linear Regression Model]{\includegraphics[scale=0.65]{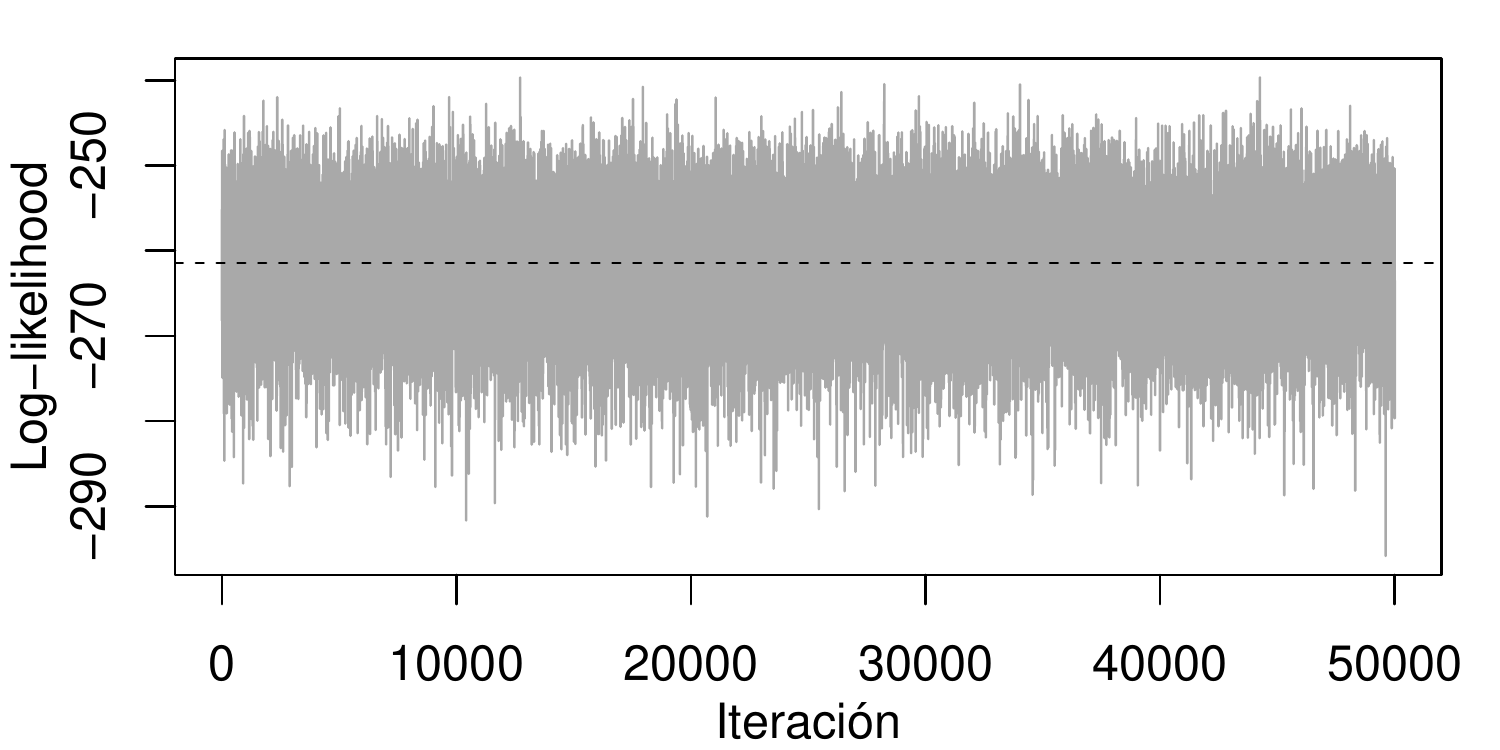}}
\caption{Log-likelihood traceplots when fitting the models for the plant size data analyzed in Section \ref{sec_application}.}
\label{fig_logliks}
\end{figure}

\section{Computation}\label{sec_computation}

We implement all the models above following the algorithms provided in the corresponding sections entitled as Posterior Inference. Every time our results are based on $B = 50,000$ samples of the posterior distribution obtained after thinning the original chains every 10 observations and a burn-in period of 10,000 iterations. In addition, before using the MCMC samples with inferential purposes, we determine first if there is any evidence of lack of convergence of any chain to its stationary distribution. Following standard practices, we produce log-likelihood traceplots of each model. Fitting the models to the data provided in Section \ref{sec_application}, such plots strongly suggest that there are no stationary concerns since the log-likelihoods move in a consistent direction (see Figure \ref{fig_logliks}). Further, autocorrelation plots of model parameters (not shown here) indicate that there are no signs of strong dependence in the chains. Thus, we are very confident of our MCMC samples to perform inductive tasks.

\section{Model Checking and Goodness-of-Fit}\label{sec_model_check}

We check the in-sample performance of the model by generating new data from the posterior predictive distribution, and then, calculating a battery of test statistics (such as the mean), aiming to compere the corresponding empirical distributions with the actual values observed in the sample \citep{gelman2013bayesian}. In this spirit, for each quantity of interest, we also compute the posterior predictive $p$-vale (ppp), which can be calculated as
$$
ppp = \textsf{Pr}(t(\yv^{\text{rep}})>t(\yv)\mid\yv)
$$
where $\yv^{\text{rep}}$ is a predictive dataset and $t$ a test statistic, in order to measure how good the model is in fitting the actual sample. 

Finally, in order to asses the goodness-of-fit of each model as a measure of their predictive performance, in what follows we consider two metrics that account for both model fit and model complexity. The goal here is not necessarily picking the model with lowest estimated prediction error but to determine if improvements in fitting the model are large enough to justify the additional difficulty. That why such measures also serve as model-selection tools. The model-based literature has largely focused on the Bayesian Information Criteria (BIC) as a mechanism for model selection.  However, the BIC is typically inappropriate for hierarchical models since the hierarchical structure implies that the effective number of parameters will typically be lower than the actual number of parameters in the likelihood \citep{gelman2014understanding}.  Two popular alternatives to the BIC that address such an issue are the Deviance Information Criterion \citep[DIC]{spiegelhalter2002bayesian,spiegelhalter-2014},
$$\DIC  = - 2\log p ( \yv \mid \hat{\mathbf{\Theta}} ) + 2p_{\DIC}\,,$$
with
$p_{\DIC} =   2\log p ( \yv \mid \hat{\mathbf{\Theta}} ) - 2\expec{\log p \left(\yv \mid \mathbf{\Theta} \right) }$,
and the Watanabe-Akaike Information Criterion \citep[WAIC]{watanabe2010asymptotic,watanabe2013widely},
$$
\WAIC  = -2\,\sum_{i,j}\log \expec{ p \left( y_{i,j} \mid \mathbf{\Theta} \right) }  + 2\,p_{\WAIC}\,,
$$
with $p_{\WAIC} = 2\sum_{i,j} \big\{ \log\expec{p \left(y_{i,j}\mid\mathbf{\Theta} \right)} - \expec{\log p \left( y_{i,j}\mid\mathbf{\Theta} \right)} \big\}$,
where $\hat{\mathbf{\Theta}}$ is the posterior mean of model parameters, and $p_{\DIC}$ and $p_{\WAIC}$ are penalty terms accounting for model complexity. Note that in the previous expressions all expectations, which are computed with respect to the posterior distribution, can be approximated by averaging over Markov chain Monte Carlo (MCMC) samples (see Section \ref{sec_computation} for details). Next section we have adopt a standard approach and use the DIC. 

\section{Illustration}\label{sec_application}

Nitrogen is an essential macronutrient deeded by all plants to thrive. It is an important component of many structural, genetic, and metabolic compounds in plant cells. Increasing the levels of nitrogen during the vegetative stage can strengthen and support the plant roots, enabling them to take in more water nutrients. This allows a plant to grow more rapidly and produce large amounts of succulent, green foliage, which in turn can generate bigger yields, tastier vegetables, and a crop that is more resistant to pets, diseases, and other adverse conditions. Using too much nitrogen, however, can be just as harmful to plants as to little. A researcher took $n=5$ measurements of nitrogen soil concentration ($x$) and plant sizes ($y$) within each of $m=24$ farms. Thus, we have that $\yij$ and $x_{i,j}$ are the plant size and the nitrogen soil concentration values, respectively, associated the $i$-th plant from the $j$-th farm, $i=1,\ldots,5$ and $j=1,\ldots,24$.

\begin{figure}[!t]
\centering
\includegraphics[scale=.68]{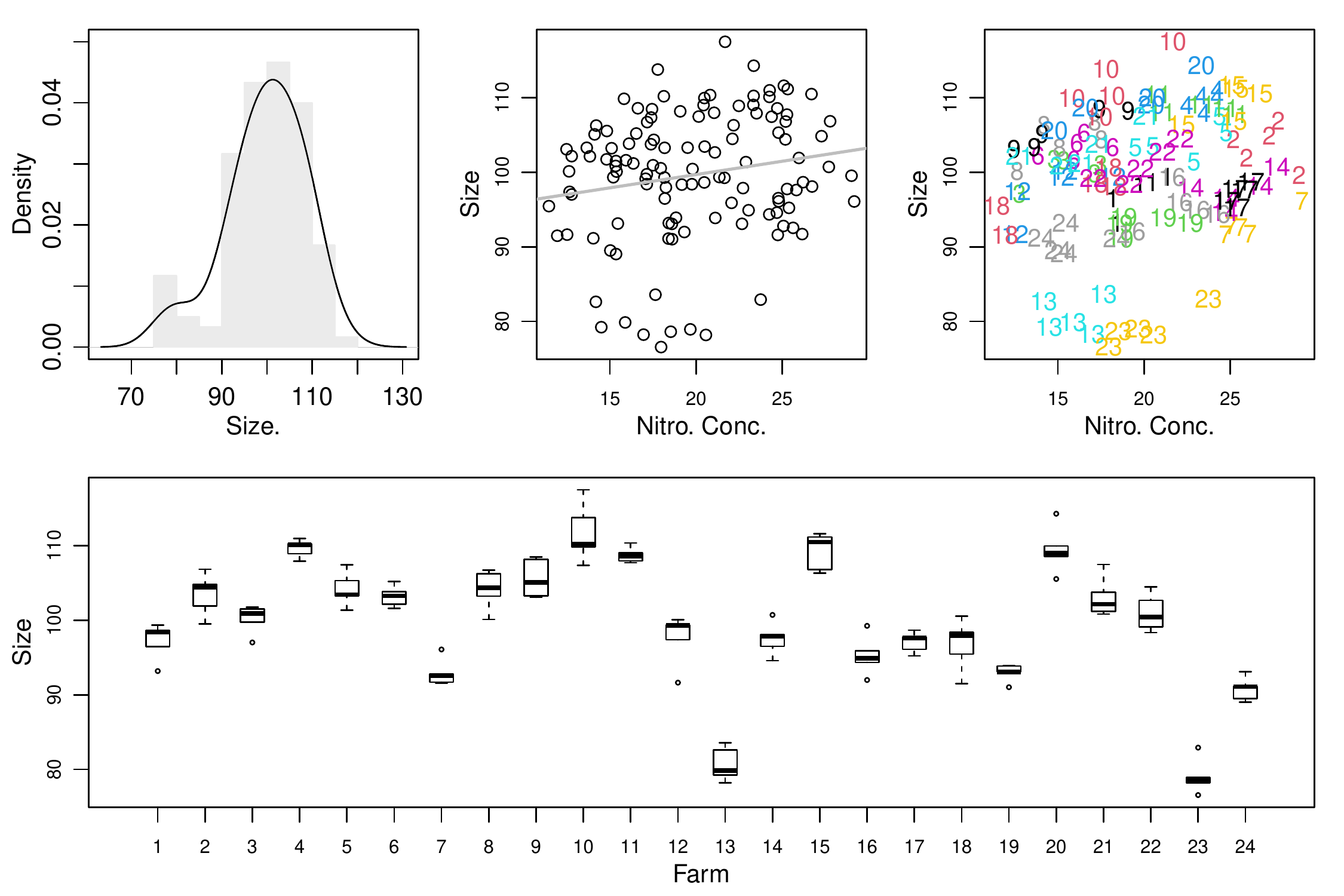}%
\caption{Descriptive plots. The second panel exhibit the ordinary least squares regression line for this data. Colors in the third panel correspond to different farms.}
\label{fig_EDA_plots}
\end{figure}

\subsection{Exploratory Data Analysis}

A histogram of the plant size is shown in the first panel of Figure \ref{fig_EDA_plots}. The plant size ranges from 76.56 to 117.50 which seems quite large in comparison with the plant size range within each farm (see the bottom panel). The second and third panels show the relationship
between the plant size and the nitrogen soil concentration. In particular, the second panel exhibits the ordinary least squares (OLS) regression line for these data, which is given by $y= 92.57 + 0.36 x$ with a standard error of $\hat{\si}_{\text{OLS}}=8.46$ with 118 degrees of freedom, and an adjusted R-squared of $R^2_{\text{adj}}= 2.56\%$. The third panel presents the same plot but taking into account the farm where the measurements belong to. These panels along with the OLS fit indicate two main features about this experiment: (1) there is an important relationship between the nitrogen soil concentration and the plant size (in our OLS fit the slope turns out to be significant, $p$-value $ = 0.04$); and (2) there is a clear farm effect on plant size since colors in the scatter plot reveal clustering patters, and also, the corresponding boxplots strongly suggest differences in terms of mean plant growth among farms.

\subsection{Normal Linear Regression Model} 

We fit the linear regression model given in Section \ref{sec_normal_linar_regression_model} to these data without taking into account the farm information. Posterior summaries of the model parameters are provided in Table \ref{tab_post_summ_MODEL_1}. Even though the posterior mean of the model parameters practically coincide with their corresponding OLS estimates, these results are again quite limited since they do not allow us to isolate any kind of effect over the plant size arising from the grouping factor. Such an impossibility strongly motivates the hierarchical approaches that we present in this paper.

\begin{table}[!h]
\centering
\begin{tabular}{ccccc}
   \hline
    Parameter  & Mean  & SD   & Q2.5\% & Q97.5\% \\ 
   \hline
    $\beta_1$  & 92.59 & 3.60 & 85.53 & 99.61 \\ 
    $\beta_2$  & 0.36  & 0.18 & 0.01  & 0.70  \\ 
    $\sigma^2$ & 72.89 & 9.67 & 56.35 & 94.07 \\ 
    \hline
\end{tabular}
\caption{Posterior summaries of the model parameters in the linear regression model.}\label{tab_post_summ_MODEL_1}
\end{table}

\subsection{Hierarchical Normal Linear Regression Model}

Now we go further and fit the hierarchical linear regression model given in Section \ref{sec_hierarchical_normal_model} to the plant size data considering both group-specific fixed effects and group-specific variances. Such a group-specif approach is very convenient because it allow us to carry out separate-group inferences as opposed to its non-hierarchical counterpart.

\begin{figure}[!t]
\centering
\includegraphics[scale=0.8]{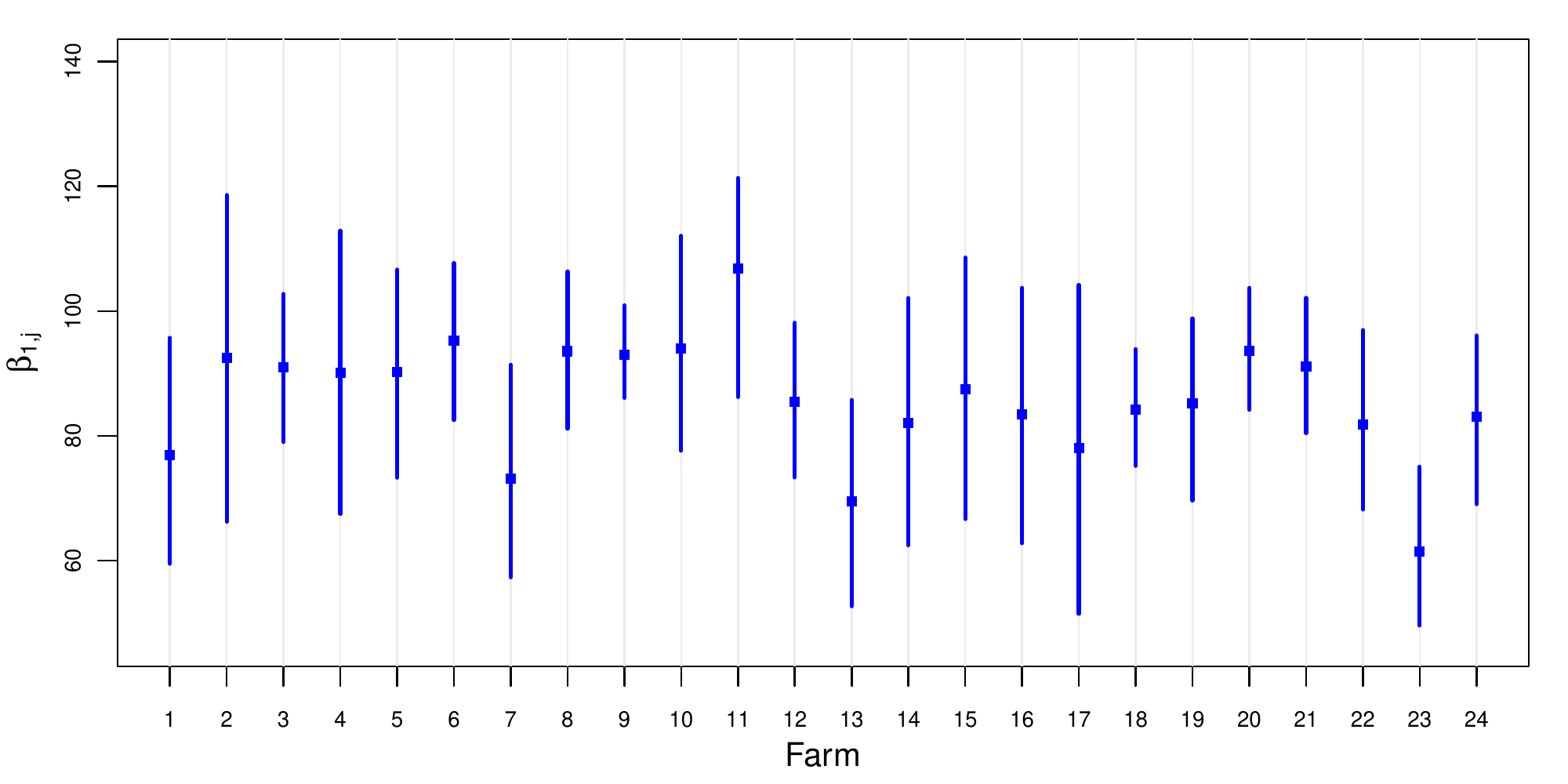}
\includegraphics[scale=0.8]{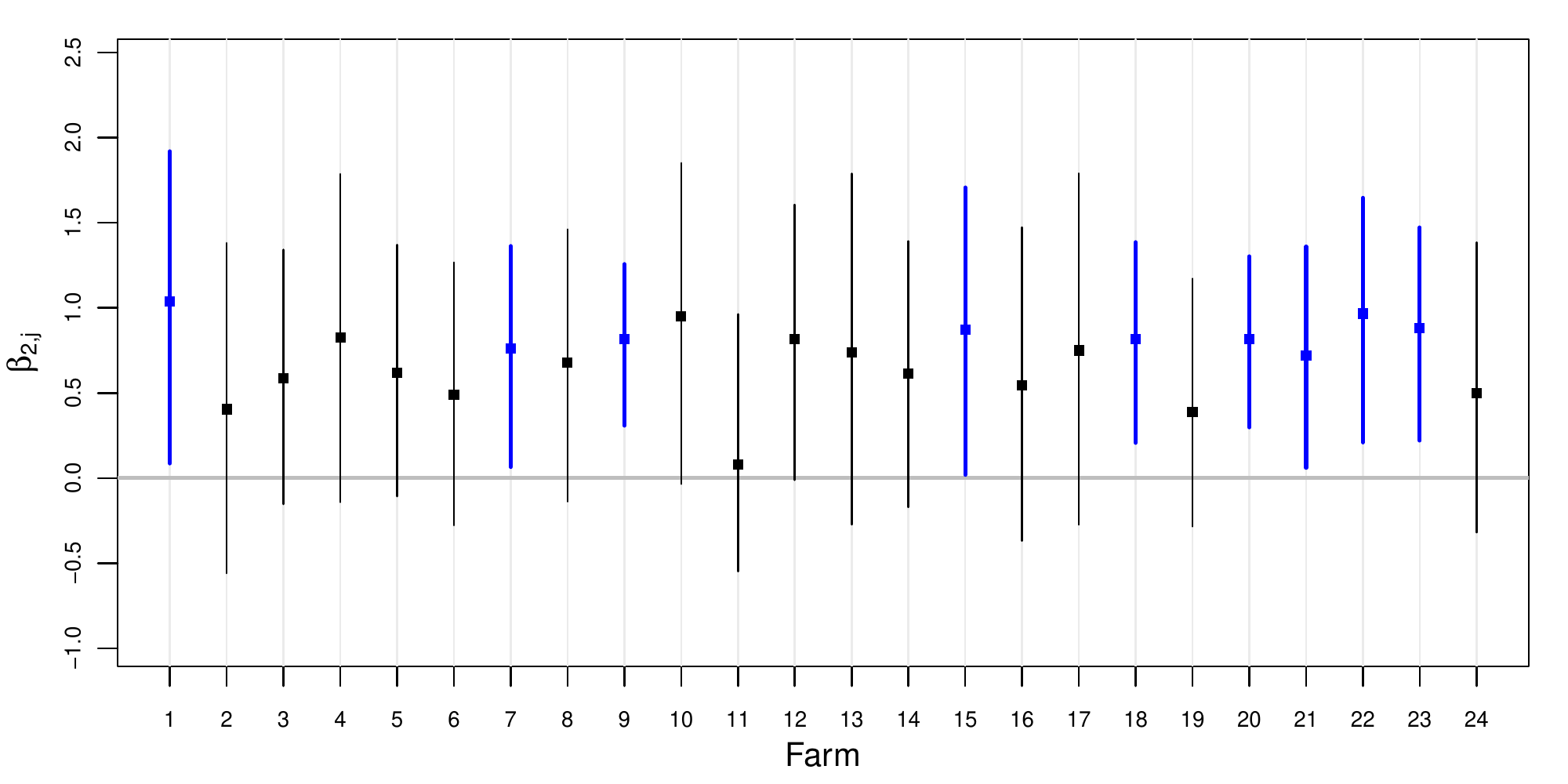}
\caption{95\% quantile-based credible intervals and posterior means (squares) for the regressor fixed-effects regressor parameters in the hierarchical Normal linear regression model. Colored thicker lines correspond to credible intervals that do not contain zero. Top panel: $\beta_{1,1},\ldots,\beta_{1,m}$ (intercepts). Bottom panel: $\beta_{2,1},\ldots,\beta_{2,m}$ (slopes).}
\label{fig_betasm2}
\end{figure}

\begin{figure}[!t]
\centering
\includegraphics[scale=0.8]{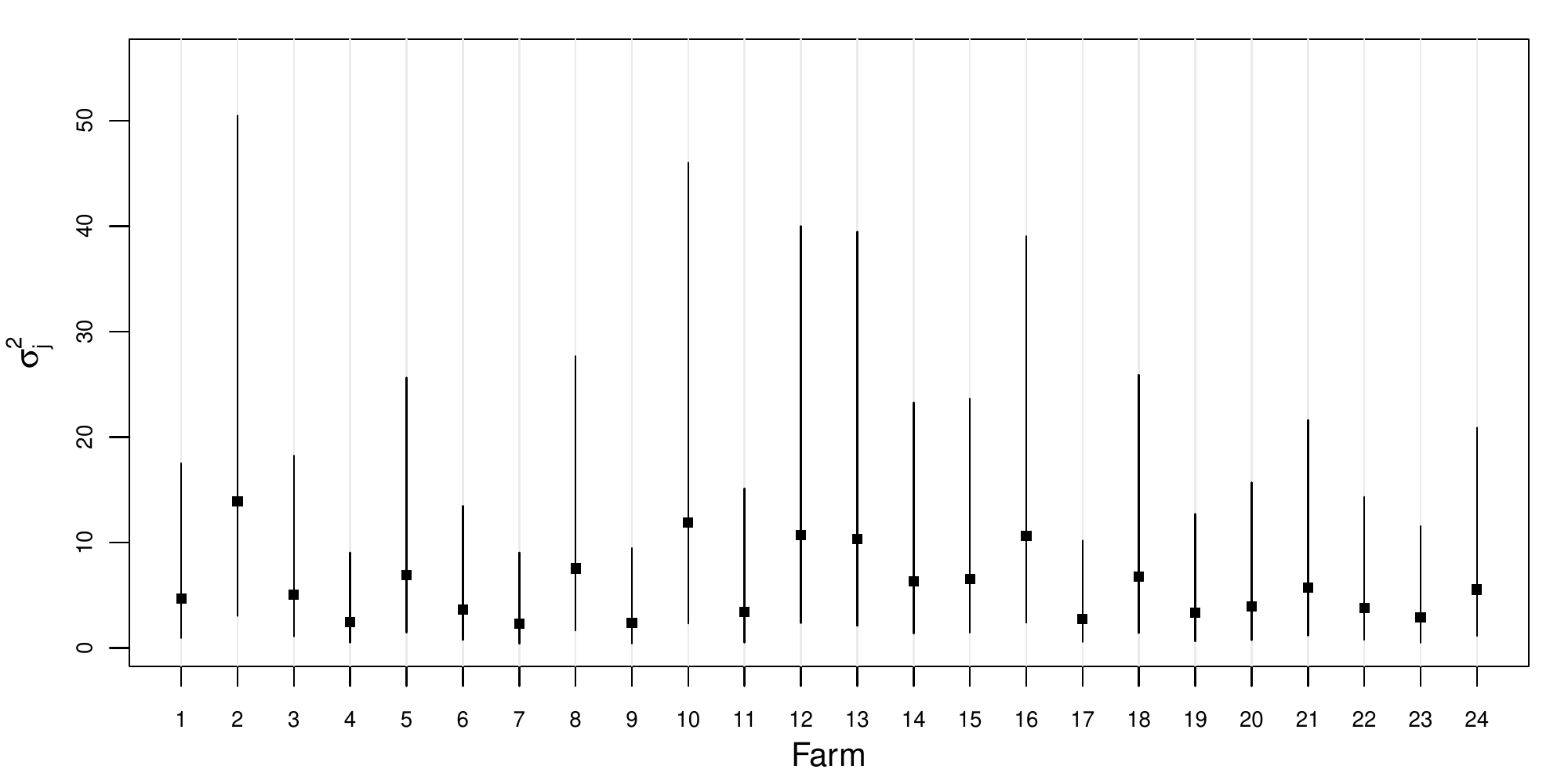}
\caption{95\% quantile-based credible intervals and posterior means (squares) for the variance parameters $\si^2_1,\ldots,\si^2_m$ in the hierarchical Normal linear regression model. }
\label{fig_sigm2}
\end{figure}

We present our main results in Figures \ref{fig_betasm2} and \ref{fig_sigm2} where we display 95\% quantile-based credible intervals for $\bev_1,\ldots,\bev_m$ and $\si^2_1,\ldots,\si^2_m$, respectively. At this point we are capable of making evident some important findings. The uncertainty about the group-specific parameters is not even since the amplitude of the credible intervals clearly varies across farms. This effect is particularly evident for the variance components. In addition, point estimates (posterior means) of the group-specific parameters are also quite variable, which strongly suggests that for these data considering this approach is beneficial because it allows us to characterize for each farm its own unique features. However, some farms show some signs of similar features, which was also evident before in Figure \ref{fig_EDA_plots}. We explore clustering patterns in the next subsection.

\begin{figure}[!t]
\centering
\includegraphics[scale=0.8]{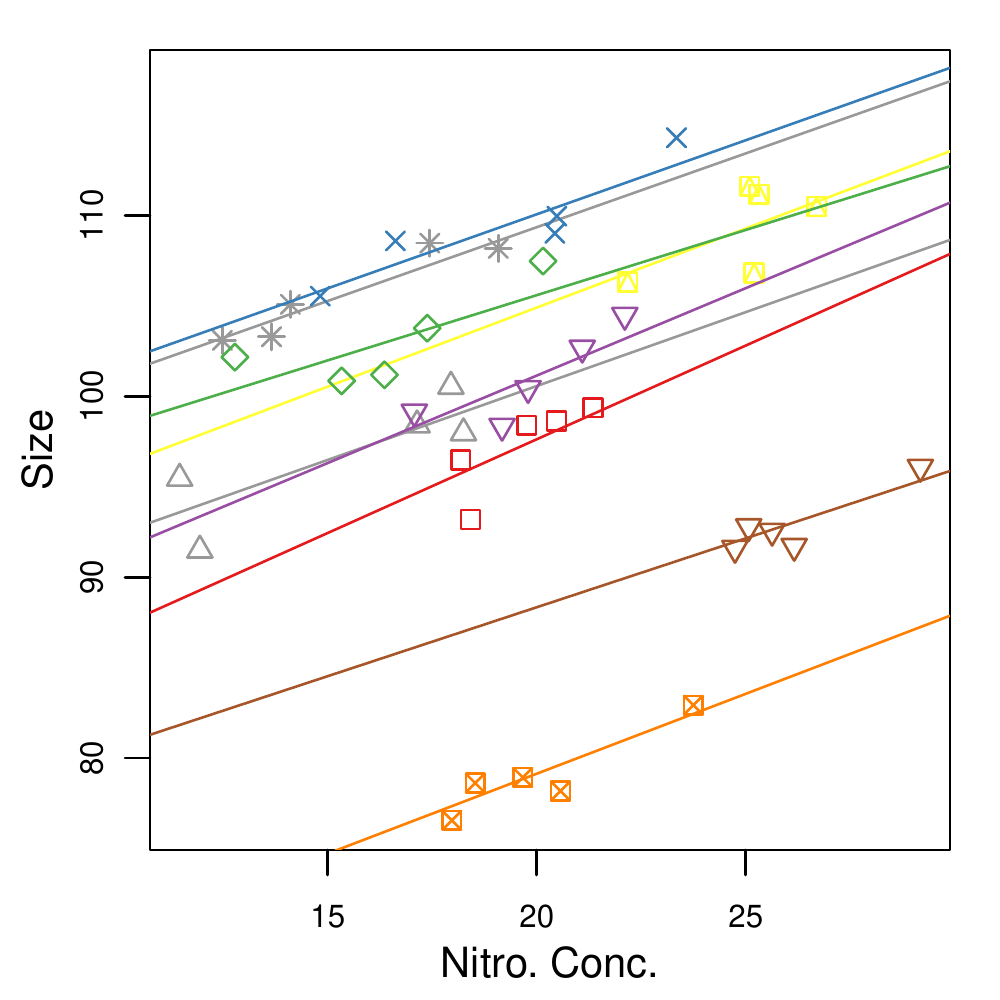}
\caption{Estimated lines of the hierarchical Normal linear regression model for those farms whose nitrogen soil concentration is statistically significant. Colors are used to represent different farms.}
\label{fig_scattersm2}
\end{figure}

As expected, we see that all the intercepts are statistically significant, but also highly variable (ranging from 61.48 to 106.81). On the other hand, the story behind the slopes (ranging from 0.08 to 1.04) is quite different. We see that just 9 out of 24 (37.5\%) of such parameters turn out to be significant, namely, for farms 1, 7, 9, 15, 18, 20, 21, 22, and 23 (Figure \ref{fig_scattersm2} shows the corresponding estimated regression lines for these farms). The previous fact strongly suggests that the relevance of the relationship between the nitrogen soil concentration and the plant size is not consistent across farms. Therefore, in this case, considering farms as a grouping factor makes a substantial impact on the analysis. 

\subsection{Clustering Hierarchical Normal Linear Regression Model}

It is quite reasonable attempting to identify clusters composed of farms given the abundant evidence of similarities among groups and cluster formation detected in both the exploratory data analysis and the hierarchical modeling stage of the analysis. Here we fit here the clustering hierarchical linear regression given in Section \ref{sec_clustering_model} model in order to provide a formal partition of farms. To this end, we fit the model using $K=m$ as a default number of communities. Such an extreme case represents the prior believe that there are no clustering patterns at all. A moderate large value of $K$ is convenient in situations like this because it allows the data to tell by itself how many non-empty clusters should be considered. We will see in what follows that out from the $m$ clusters, many turn out to be empty and only a few remain. 

\begin{figure}[!t]
\centering
\subfigure[Posterior distr. of $K^*$.]{\includegraphics[scale=0.7]{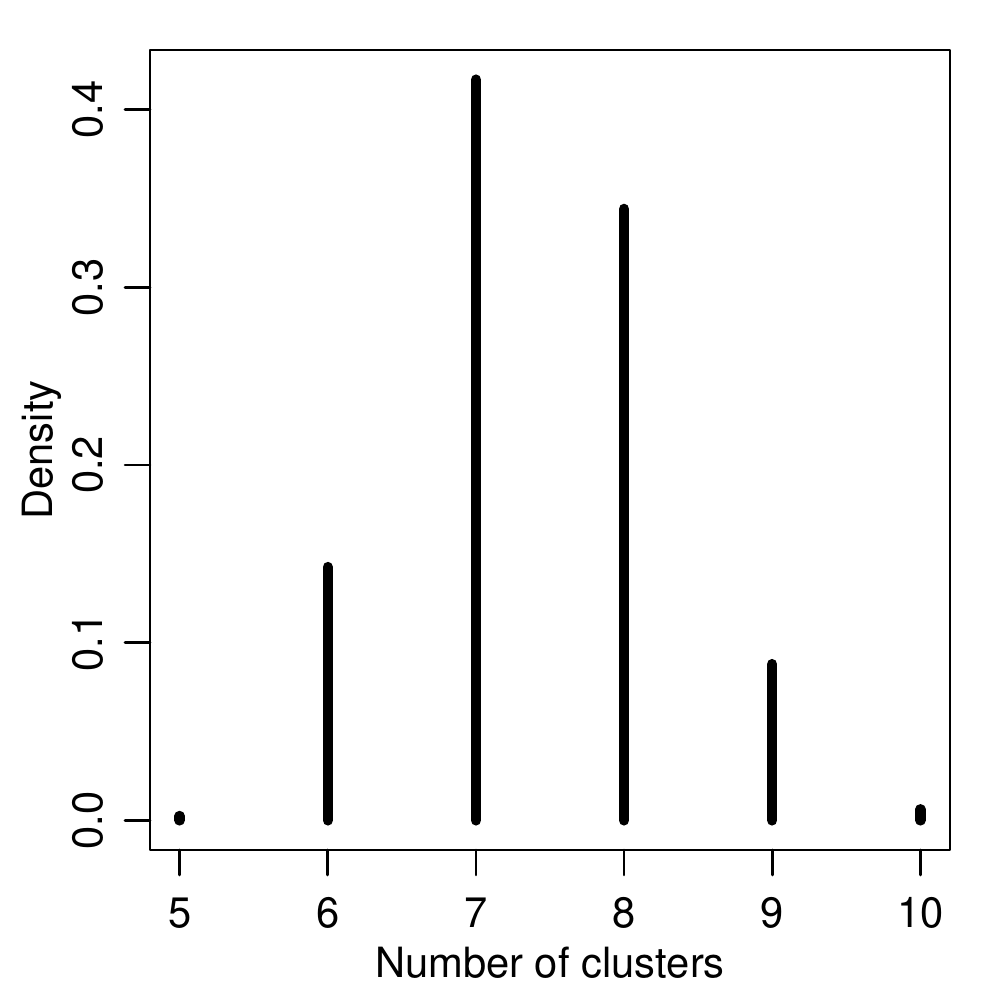}}
\subfigure[Incidence matrix $\Am$.]{\includegraphics[scale=0.7]{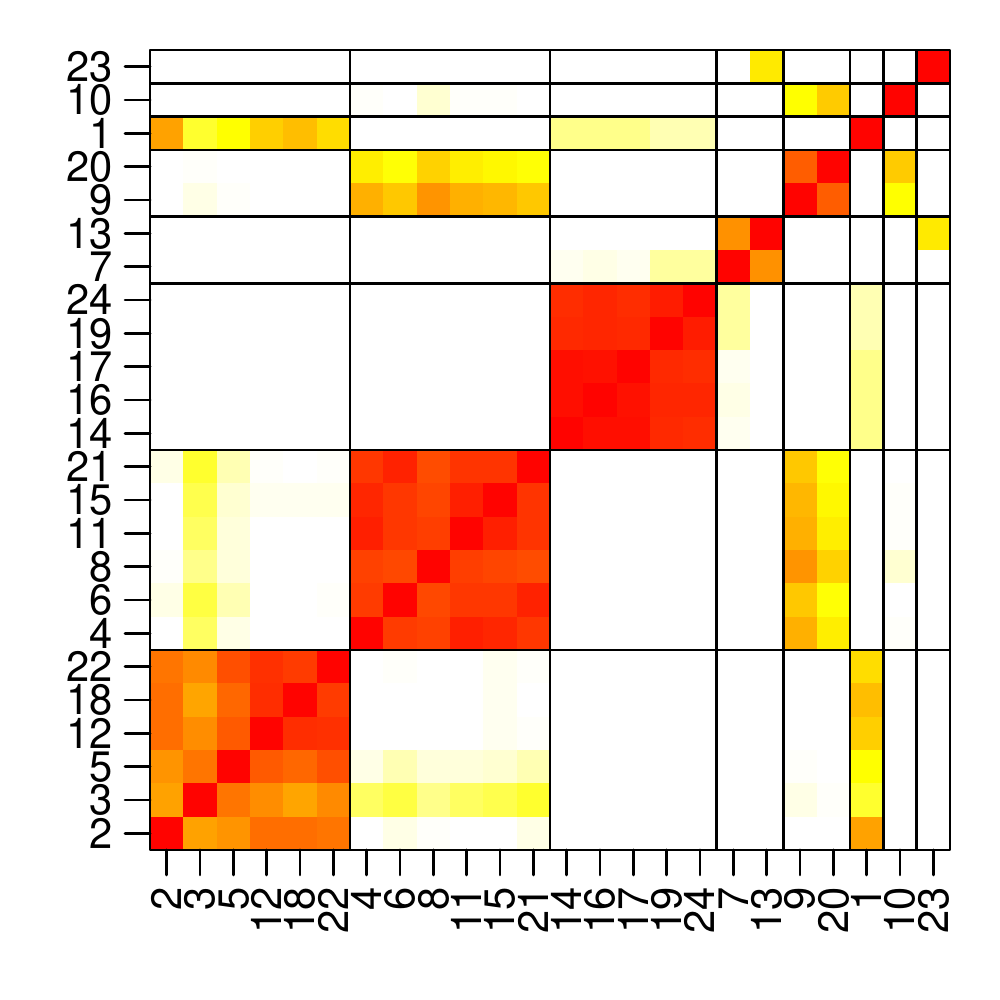}\includegraphics[scale=0.355]{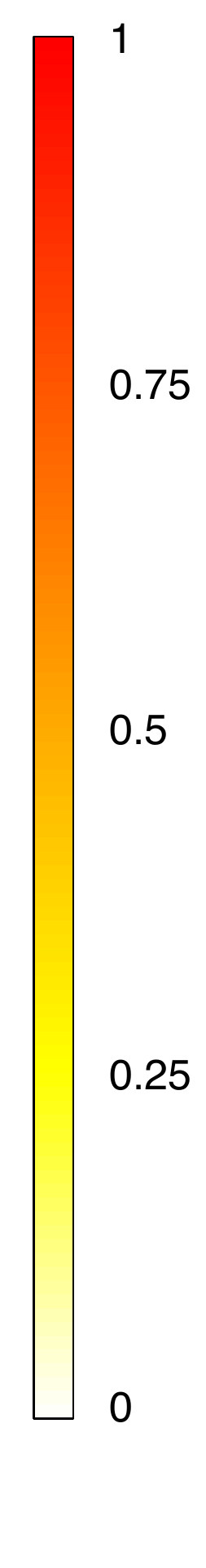}}
\caption{Posterior inference on the cluster asingments $\ga_1,\ldots,\ga_m$.}
\label{fig_incidence}
\end{figure}

Let $K^*$ be the number of non-empty clusters in the partition induced by $\ga_1,\ldots,\ga_m$. The left panel of Figure \ref{fig_incidence} shows the posterior distribution of $K^*$. We see that the most highly probable values are $K^*=7$ and $K^*=8$. In fact, $\textsf{Pr}(K^*\in\{7,8\}\mid\yv,\Xm) = 0.76$, which means that around three quarters of the posterior partitions are composed of either 7 or 8 clusters of farms. The estimated number of non-empty clusters in the data is obviously $K^*=7$ (maximum a posteriori with a reference value very close to 0.4). 

\begin{figure}[!t]
\centering
\subfigure[Cluster 1.]{\includegraphics[scale=0.37]{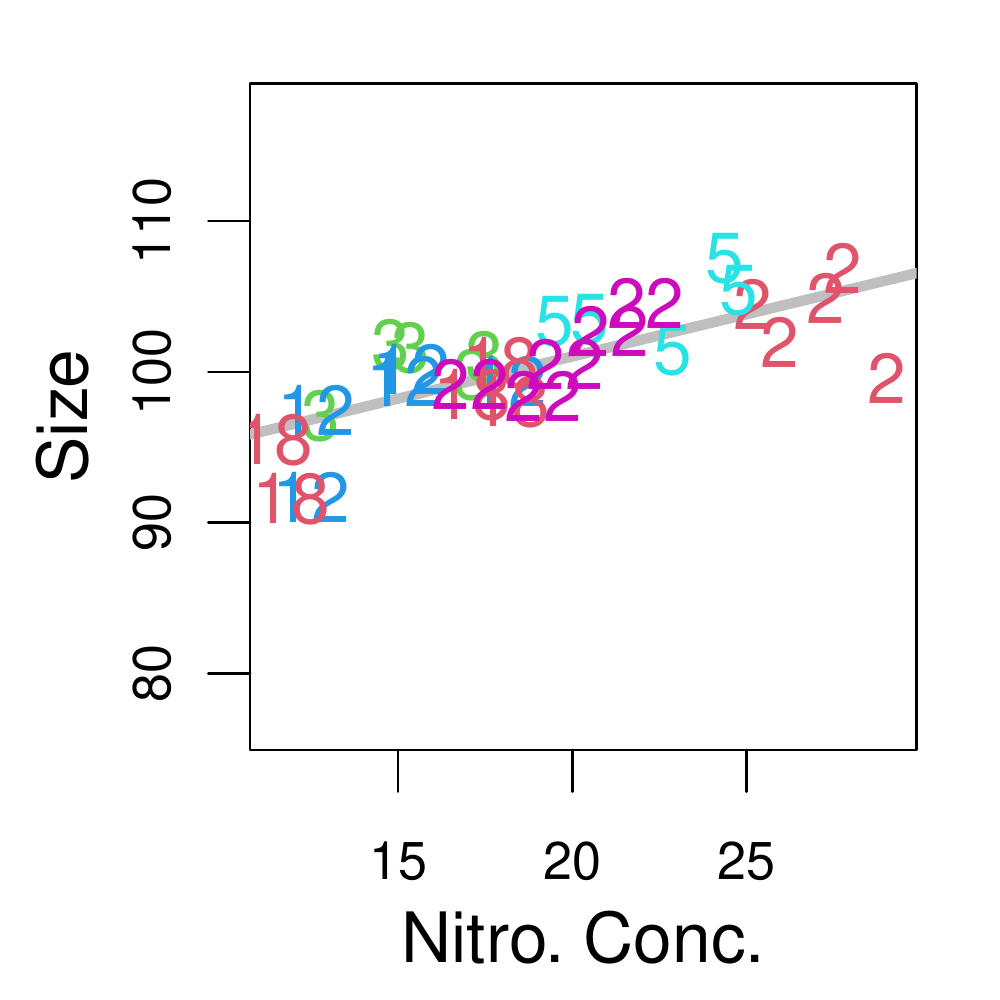}}
\subfigure[Cluster 2.]{\includegraphics[scale=0.37]{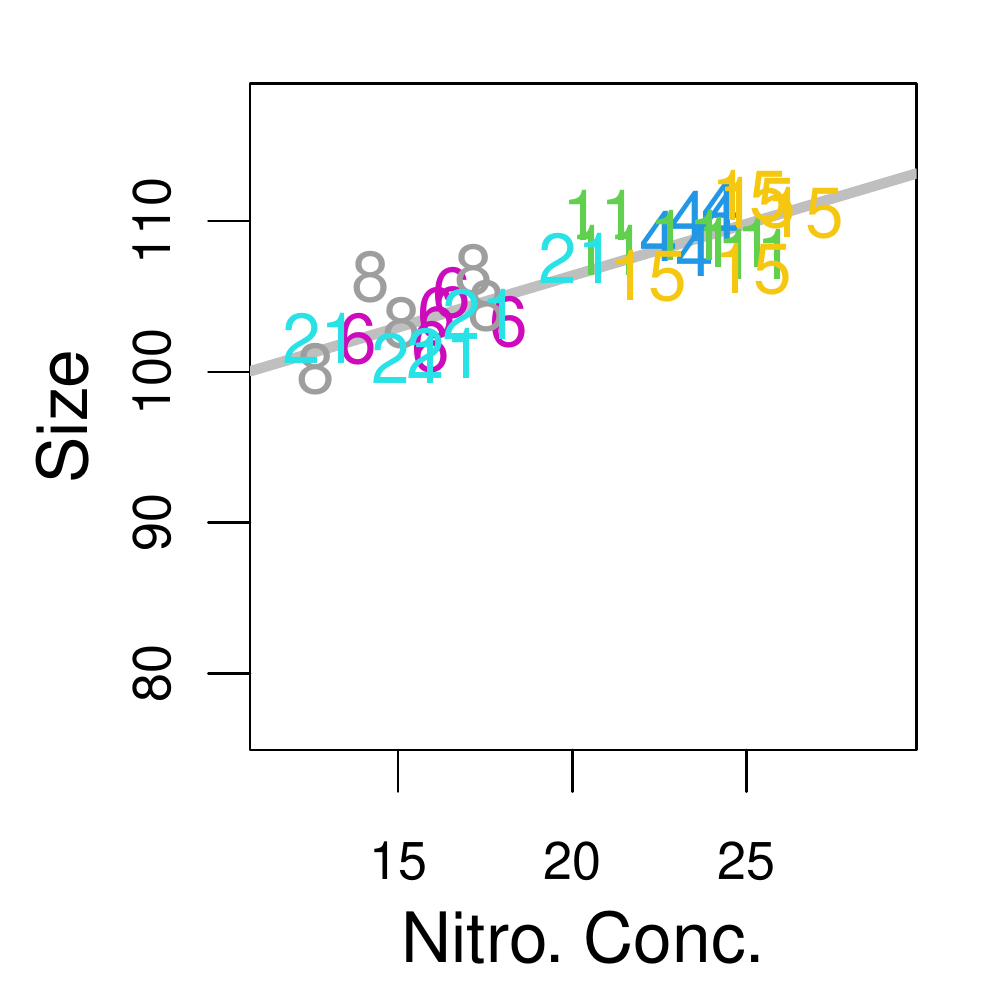}}
\subfigure[Cluster 3.]{\includegraphics[scale=0.37]{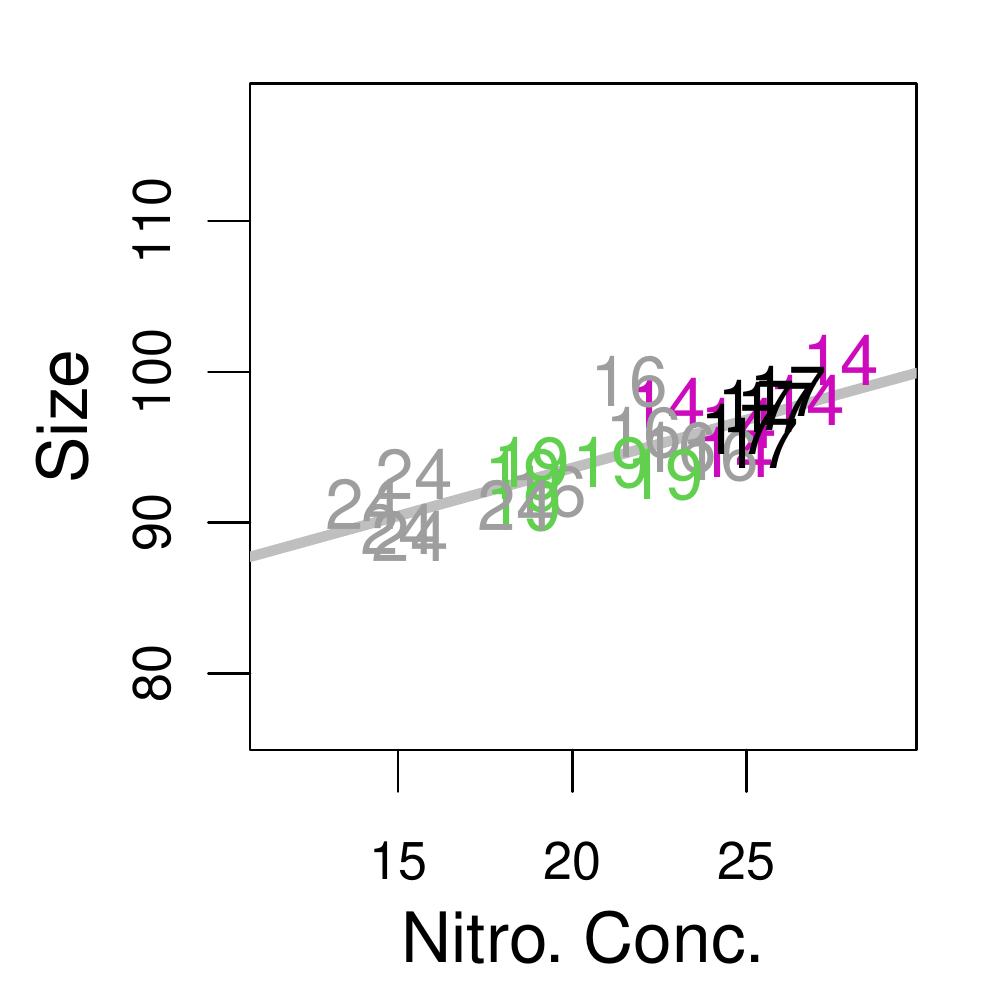}}
\subfigure[Cluster 4.]{\includegraphics[scale=0.37]{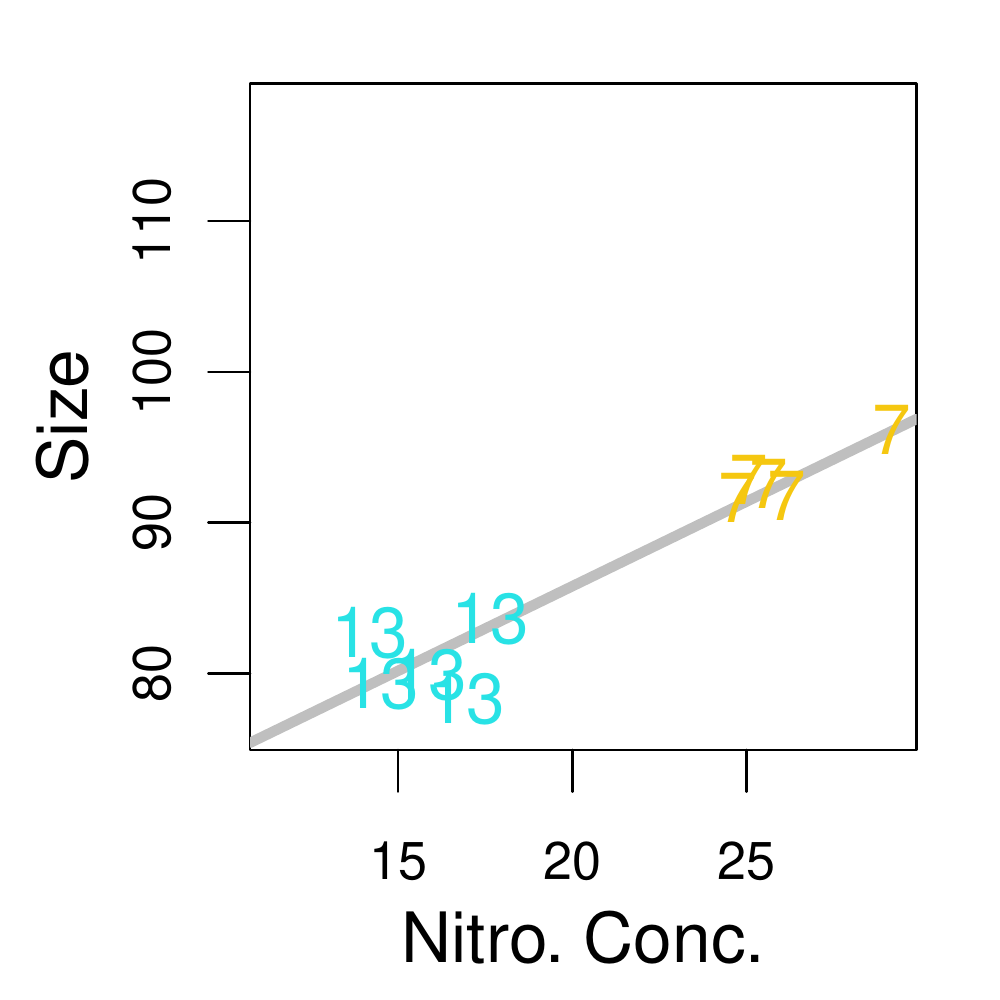}}
\subfigure[Cluster 5.]{\includegraphics[scale=0.37]{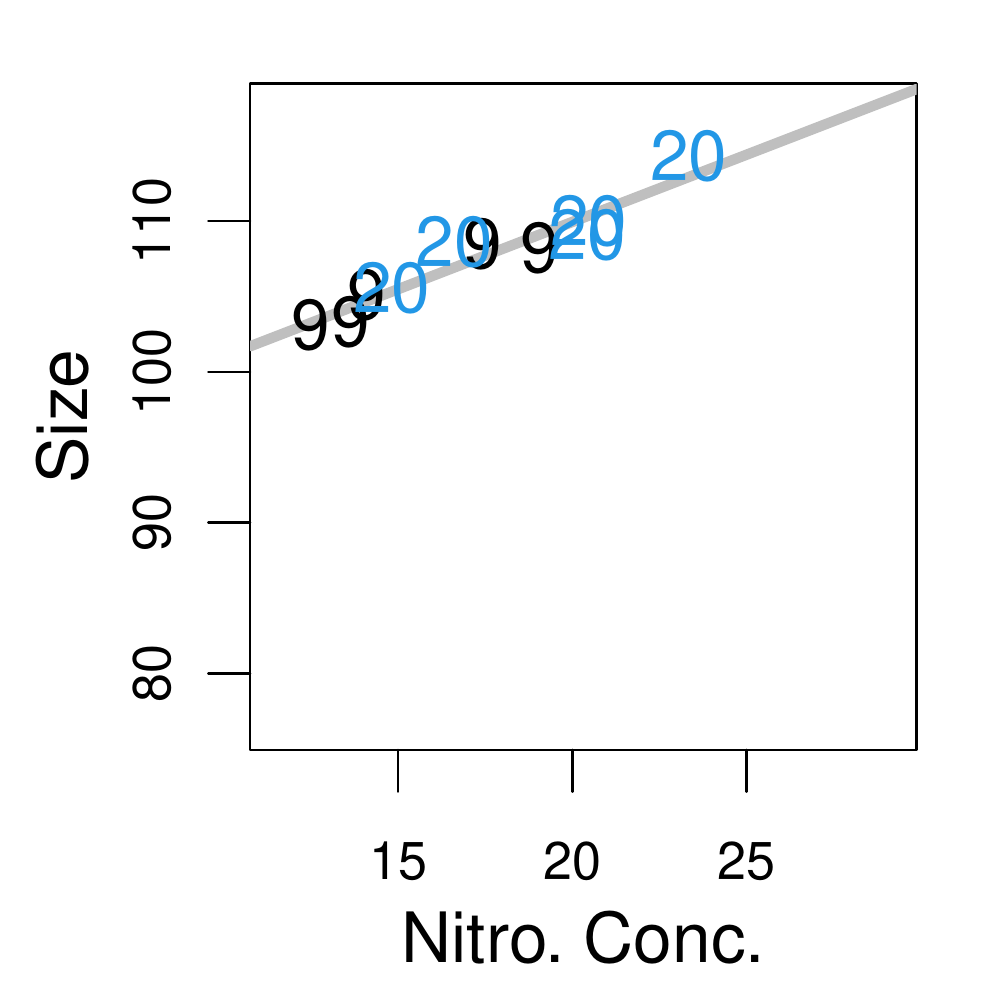}}
\subfigure[Cluster 6.]{\includegraphics[scale=0.37]{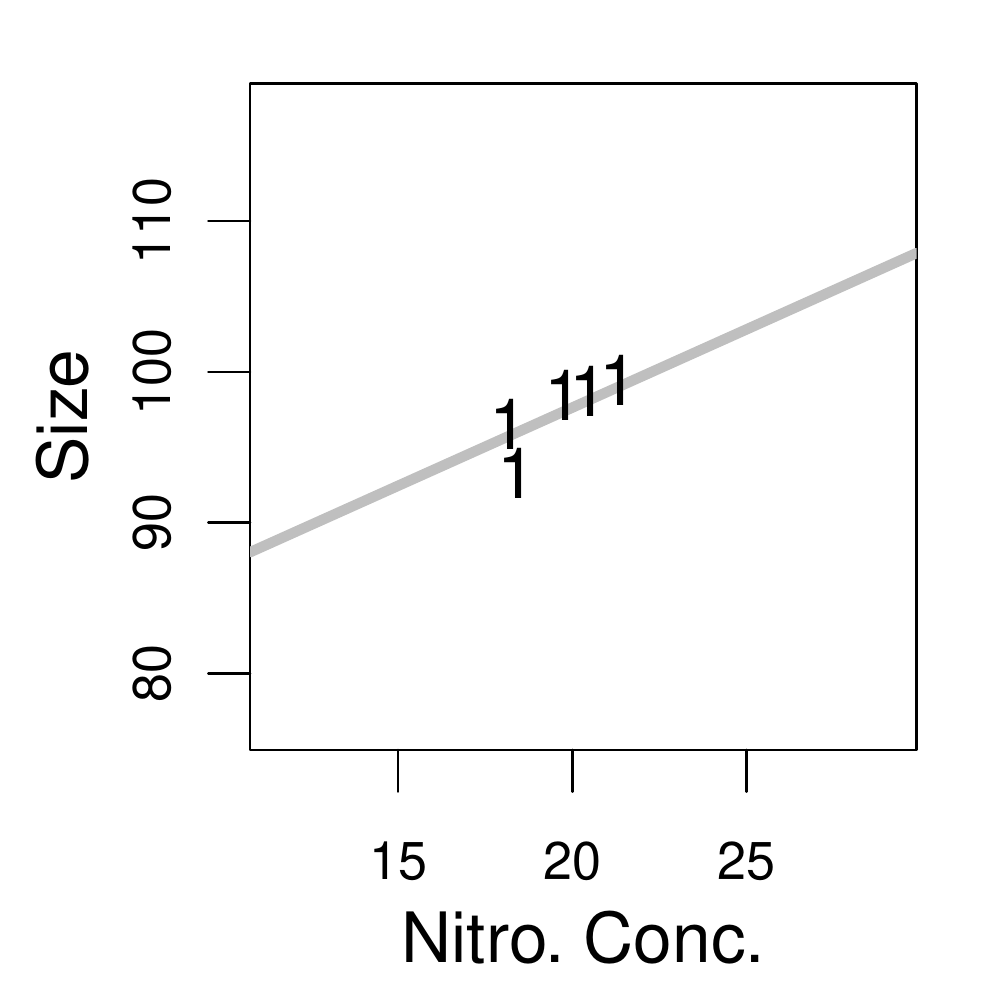}}
\subfigure[Cluster 7.]{\includegraphics[scale=0.37]{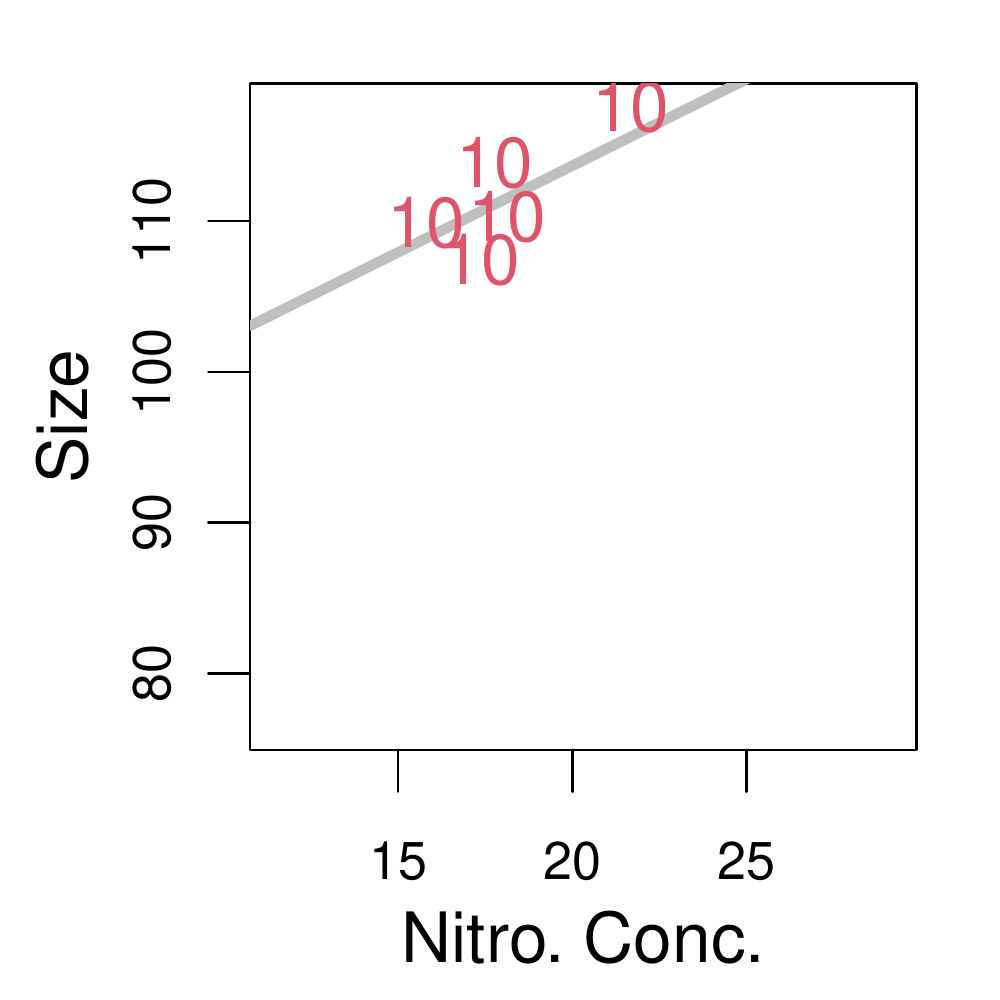}}
\subfigure[Cluster 8.]{\includegraphics[scale=0.37]{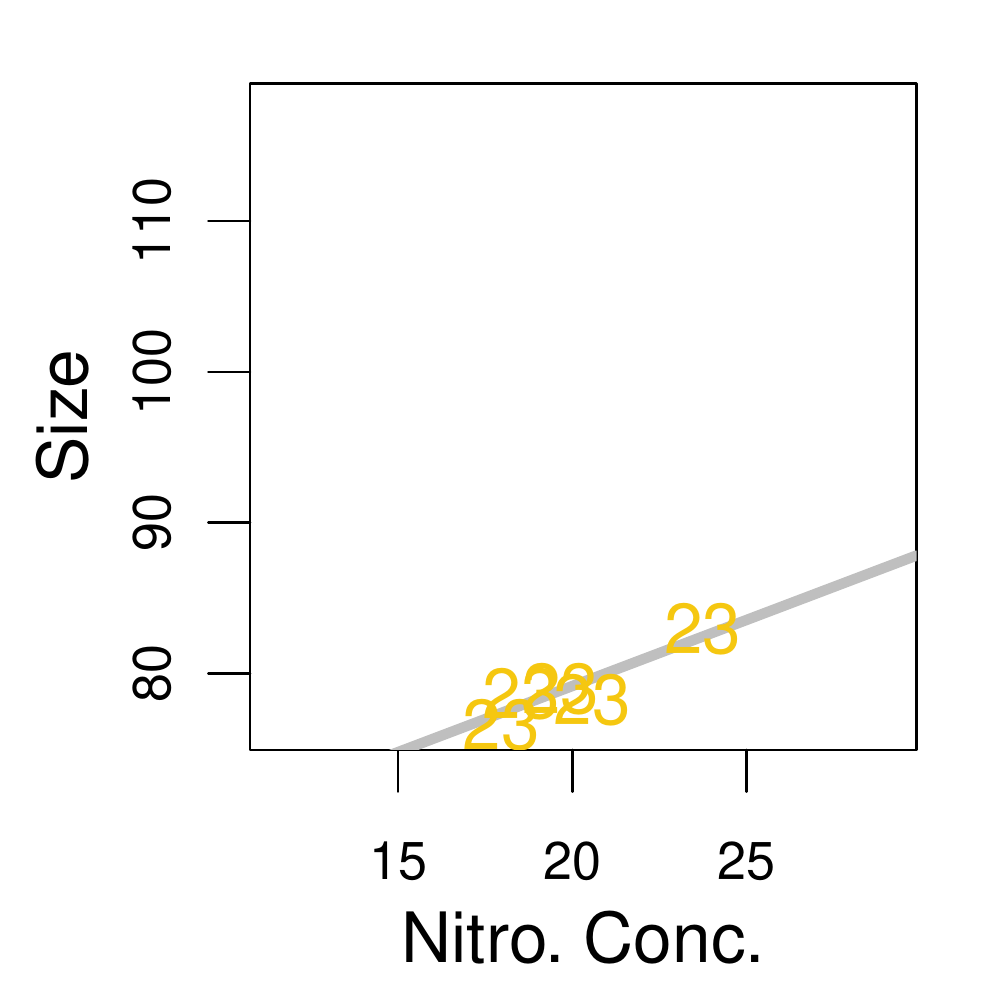}}
\caption{Scatterplots along with their estimated regression lines associated with the clusters in the partition. Colors represent represent farms. }
\label{fig_scatterclusts}
\end{figure}

On the other hand, the right panel in Figure \ref{fig_incidence} shows the $m\times m$ incidence matrix $\Am=[a_{j,j'}]$ obtained from the posterior distribution of the cluster assignments $\ga_1,\ldots,\ga_m$. The incidence matrix is a pairwise-probability matrix whose elements are given by $a_{j,j'} = \textsf{Pr}(\gamma_{j} = \gamma_{j'}\mid\yv,\Xm)$, for $j,j'=1,\ldots,m$ (note that $a_{j,j}=1$ for all $j$). Thus, $a_{j,j'}$ simply represents the posterior probability that farms $j$ and $j'$ belong to the same community. Such probabilities are indeed identifiable, however labels themselves are not since the likelihood is invariant to relabelling of the mixture components (this is known as the label switching problem; see \citealp{stephens2000dealing} and references therein). On top the incidence matrix, we also present a point estimate of the partition induced by such a matrix (represented by black lines), which can be obtained by employing the clustering methodology proposed in \cite{lau2007bayesian} with a relative error cost of 0.5. As expected, we see that eight clusters are formed from the data. The corresponding cluster sizes are 6, 6, 5, 2, 2, 1, 1, and 1. Specifically, the clusters are composed of the following farms: $C_1=\{2,3,5,12,18,22\}$, $C_2=\{4,6,8,11,15,21\}$, and $C_3=\{14,16,17,19,24\}$, $C_4=\{7,13\}$, $C_5=\{9,20\}$, $C_6=\{1\}$, $C_7=\{10\}$, and $C_8=\{23\}$. These clusters along with their estimated regression lines are also represented in Figure \ref{fig_scatterclusts}. We see that the estimated clusters make sense spatially according to the data points. The slope parameters turn out    to be significant for every cluster. 

\subsection{Model Checking and Goodness-of-Fit}

First, we evaluate in-sample predictive performance of each model by means of the mean square error (MSE) of replicated data as well as the posteriori predictive $p$-values (ppp's) associated with the predictive distribution of a set of test statistics (mean, median, interquartile range, and the standard deviation), both locally and globally (i.e., with and without considering the farm as grouping factor, respectively). Furthermore, as discussed in Section \ref{sec_model_check}, we also consider the deviance information criterion (DIC) in order to assess the overall goodness-of-fit of the models.

In what follows we examine the performance of all the fitted models, namely, the linear regression model (LRM), the hierarchical linear regression model (HLRM), and the clustering hierarchical linear regression model (CHLRM). Our main findings at a global level are presented in Table \ref{tab_check}. As expected, the performance of LRM at predicting new data is the worst. Interestingly, both HLRM and CHLRM practically have the same behavior in this regard. A similar conduct is encountered again in terms of model fit, but this time the DIC tends to favor HLRM over CHLRM. These results strongly suggest that considering a hierarchical structure when building a model clearly favor both in-sample predictive performance as well as goodness-of-fit. 

\begin{table}[!h]
\centering
\begin{tabular}{cccc}
   \hline
    Model & MSE & $p_\text{DIC}$ & DIC \\ 
   \hline
    LRM   & 144.528 & 2.972  & 857.158 \\ 
    HLRM  & 10.206  & 30.469 & 526.450 \\ 
    CHLRM & 10.068  & 32.167 & 555.020 \\ 
    \hline
\end{tabular}
\caption{Global measures associated with the mean square error, posterior predictive $p$-values, and deviance information criterion for all the models fitted to the plant size data. LRM: Normal linear regression model. HLRM: Hierarchical Normal linear regression model. CHLRM: Clustering hierarchical Linear regression model. }\label{tab_check}
\end{table}

\begin{figure}[!t]
\centering
\subfigure[LRM.]  {\includegraphics[scale=0.5]{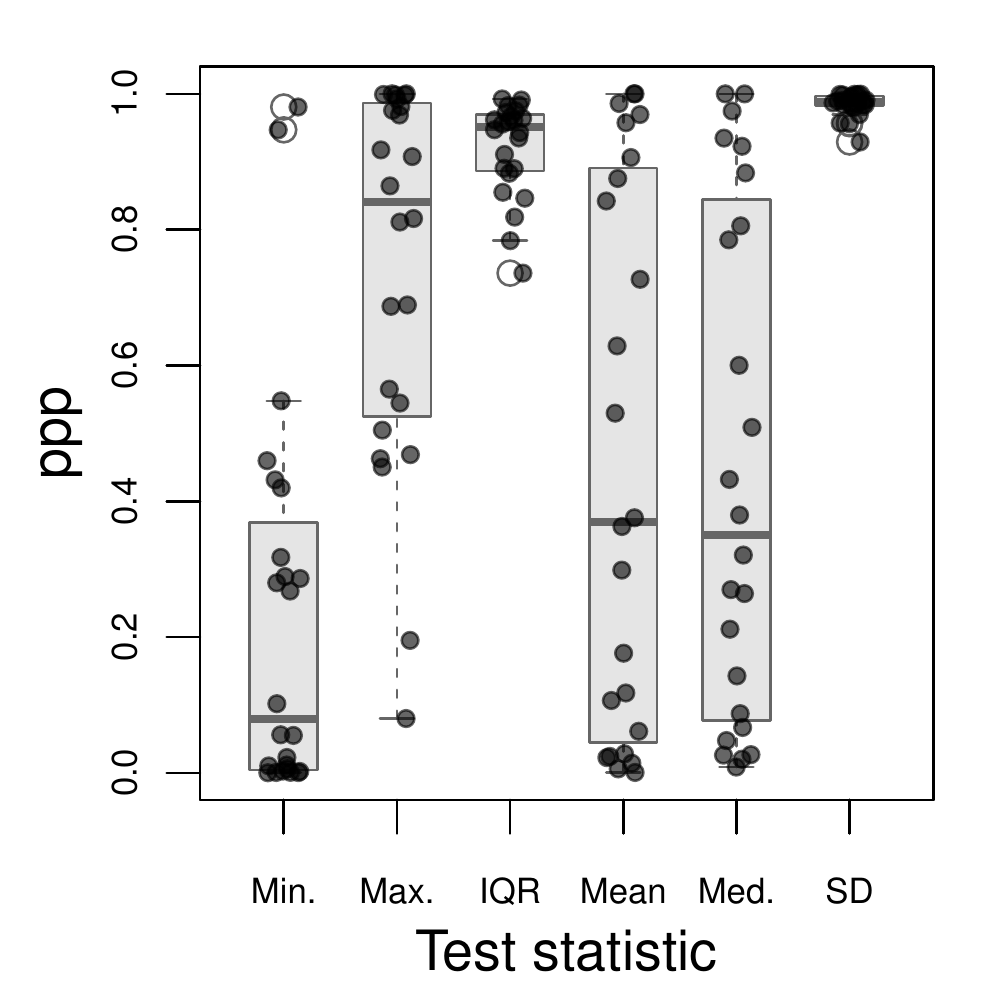}}
\subfigure[HLRM.] {\includegraphics[scale=0.5]{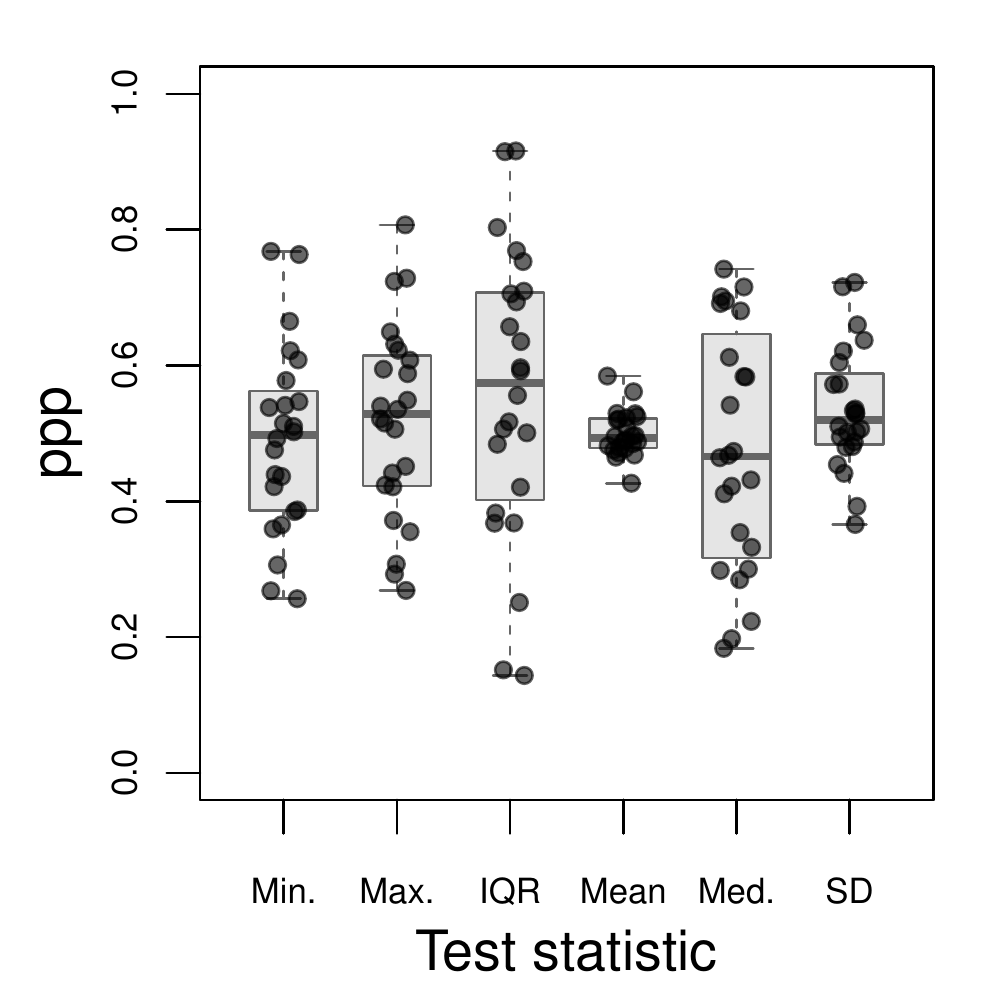}}
\subfigure[CHLRM.]{\includegraphics[scale=0.5]{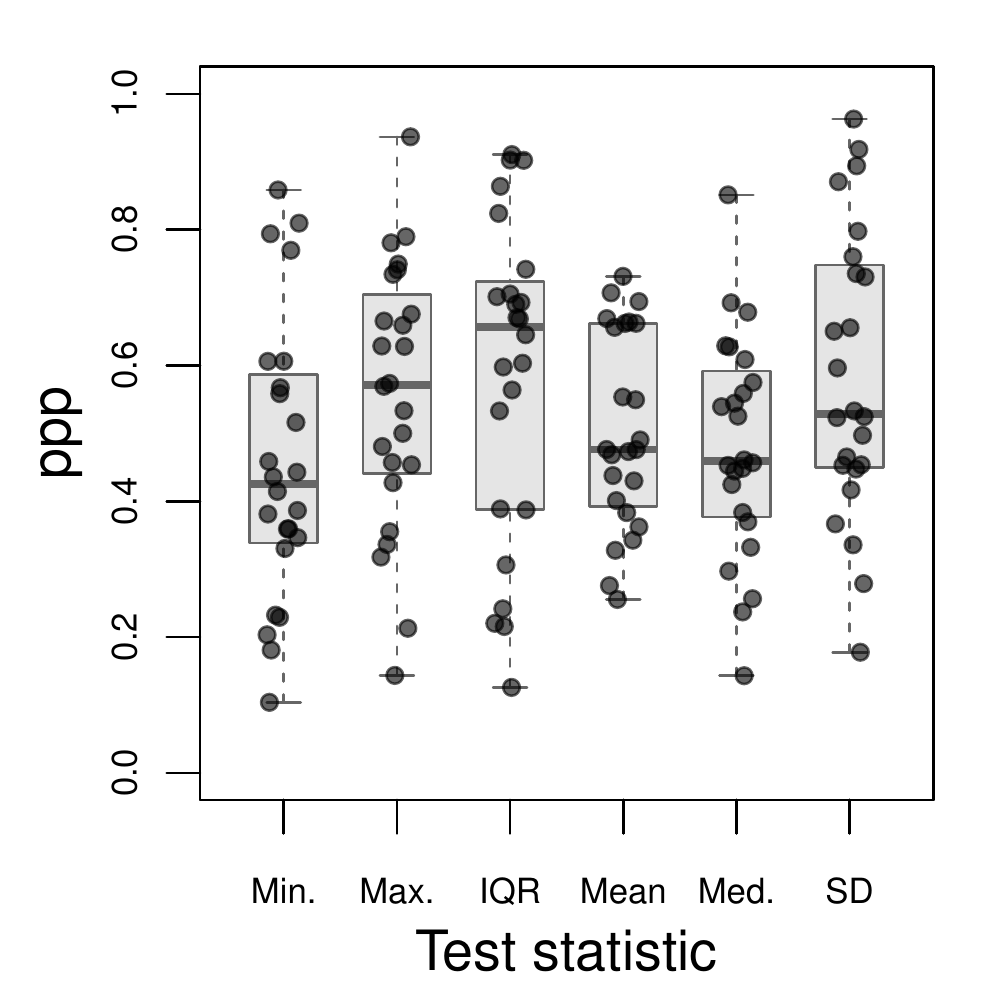}}
\caption{Local ppp's for a battery of test statistics.}
\label{fig_ppp}
\end{figure}

At global level, all the models seem to predict adequately the test statistics since there are no evidence of extreme ppp values close to either 0 or 1 (table not shown here). However, the story locally is quite different. Figure \ref{fig_ppp} show boxplots summarizing the ppp distribution at a local level (i.e., within each farm). As opposed to its early global conduct, LRM misbehaves and fails at capturing the the test statistics because the ppp's are too extreme. On the other hand, both HLRM and CHLRM fit the data properly since the ppp distribution is approximately centered around 0.5. However, the ppp's for HLRM are less spread than those for CHLRM, which indicates that HLRM tends to have a mild improvement in predictive performance in this case.  

\section{Discussion} \label{sec_discussion}

Hierarchical models provide a strong alternative to analyze
complex and realistic settings. Their parameter flexibility allow us to describe many characteristics of a given dataset that a regular single-level model does not provide. The ability to model within and between means and variances yields to better knowledge of the problem (even if we want to predict future values at any stage). For even more detail and deep types of complexity, this kind of models provide many alternatives for generalizations as discussed in Section \ref{sec_extensions}.

If additional time-dependent covariates were available, we would be able to extend the linear dependency in the model by letting $\textsf{E}(y(t)\mid\xv,\bev(t))=\xv^{\textsf{T}}\bev(t)$,
where $\bev(t) = (\beta_{1}(t),\ldots,\beta_{d}(t))$
is a vector of arbitrary real smooth functions called dynamic parameters. This model plays a fundamental role identifying and characterizing dynamic tendencies and patterns over time in many scientific areas, such as biology, epidemiology and medical science, among others \citep{sosa2021time}.

Finally, we recommend consider alternative inference methods in order to account for larger datasets, which is currently an active research area in computational statistics (e.g., \citealp{blei2017variational})

%\nocite{*}
\bibliography{references}
\bibliographystyle{apalike}

\appendix

\section{Dirichlet process}

A random distribution function $F$ is generated from a Dirichlet Process with parameters $\alpha>0$ and $G$ a distribution function on $\mathbb{R}$, denoted by $F\sim\textsf{DP}(\alpha,G)$, if for any finite measurable partition $B_1,\ldots,B_k$ of $\mathbb{R}$,
$$
(F(B_1),\ldots,F(B_k))\sim\textsf{Dir}(\alpha G(B_1),\ldots,\alpha G(B_k))\,.
$$
$G$ plays the role of the center of the DP (also referred to as base
probability measure, or base distribution), where as $\alpha$ can be viewed as a precision parameter (the larger $\alpha$ is, the closer we expect a
realization $F$ from the process to be to $G$). See \cite{ferguson1973bayesian} for the role of $G$ on more technical properties of the DP.

Alternatively, the constructive definition of the DP \citep{sethuraman1994constructive} states that $F\sim\textsf{DP}(\alpha,G)$ if $F$ is (almost surely) of the form
$$
F(\cdot) = \sum_{k=1}^{\infty}\omega_k\delta_{\vartheta_{k}}(\cdot)
$$
where $\delta_\vartheta(\cdot)$ denotes a point mass at $\vartheta$ (degenerate distribution putting probability one on $\vartheta$), 
$\vartheta_k\simiid G$, $\omega_k=z_k\prod_{\ell=1}^{k-1}(1-z_\ell)$, $z_k\simiid\textsf{Beta}(1,\alpha)$, for $k=1,2,\ldots$.
Hence, the DP generates distributions that can be represented as countable
mixtures of point masses (the locations $\vartheta_k$ arise i.i.d. from the base distribution $G$), whose weights $\omega_k$ arise through a stick-breaking construction (it can be shown that $\sum_{k=1}^\infty\omega_k=1$ almost surely).

Based on its constructive definition, it is evident that the DP generates
(almost surely) discrete distributions on $\mathbb{R}$.

\section{Notation}

The cardinality of a set $A$ is denoted by $|A|$. If P is a logical proposition, then $1_{\text{P}} = 1$ if P is true, and $1_{\text{P}} = 0$ if P is false. $\floor{x}$ denotes the floor of $x$, whereas $[n]$ denotes the set of all integers from 1 to $n$, i.e., $\{1,\ldots,n\}$. The Gamma function is given by $\Gamma(x)=\int_0^\infty u^{x-1}\,e^{-u}\,\text{d}u$. 

Matrices and vectors with entries consisting of subscripted variables are denoted by a boldfaced version of the letter for that variable. For example, $\xv = (x_1,\ldots, x_n)$ denotes an $n\times1$ column vector with entries $x_1,\ldots, x_n$. We use $\zerov$ and $\boldsymbol{1}$ to denote the column vector with all entries equal to 0 and 1, respectively, and $\Ima$ to denote the identity matrix. A subindex in this context refers to the corresponding dimension; for instance, $\Ima_n$ denotes the $n\times n$ identity matrix. The transpose of a vector $\xv$ is denoted by $\xv^\trans$; analogously for matrices. Moreover, if $\Xm$ is a square matrix, we use $\text{tr}(\Xm)$ to denote its trace and $\Xm^{-1}$ to denote its inverse. The norm of $\xv$, given by $\sqrt{\xv^\trans\xv}$, is denoted by $\|\,\xv\|\,$.

Now, we present the form of some standard probability distributions:
\begin{itemize}

    \item Multivariate normal:
    
    A $d\times 1$ random vector $\Xv=(X_1\ldots,X_d)$ has a multivariate Normal distribution with parameters $\muv$ and $\Sig$, denoted by $\Xv\mid\muv,\Sig \sim \textsf{N}_d(\muv,\Sig)$, if its density function is
    $$
    p(\xv\mid\muv,\Sig) = (2\pi)^{-d/2}\,|\Sig|^{-1/2}\,\exp{\left\{-\tfrac12 (\xv - \muv)^{\textsf{T}}\Sig^{-1}(\xv - \muv) \right\}}\,.
    $$
    
    \item Inverse Wishart:
    
    A $d\times d$ random matrix $\mathbf{W}$ has a Inverse Wishart distribution with parameters $\nu$ y $\Sm^{-1}$, i.e., $\mathbf{W} \sim \textsf{WI}(\nu, \Sm^{-1})$, if its density function is
    $$
    p(\mathbf{W}) \propto |\mathbf{W}|^{-(\nu+d+1)/2}\,\exp{\left\{-\tfrac12\text{tr}(\Sm\mathbf{W}^{-1}) \right\}}\,,\quad \nu>0\,,\quad \Sm > 0. 
    $$

    \item Gamma:
    
    A random variable $X$ has a Gamma distribution with
    parameters $\al,\be > 0$, denoted by $X\mid\al,\be\sim\textsf{G}(\al,\be)$, if its density function is
    $$
    p(x\mid\al,\be) = \frac{\be^\al}{\Gamma(\al)}\,x^{\al-1}\,\exp{\{-\be x\}}\,,\quad x>0\,.
    $$
    
    \item Inverse Gamma:
    
    A random variable $X$ has an Inverse Gamma distribution with parameters $\al,\be > 0$, denoted by $X\mid\al,\be\sim\textsf{IG}(\al,\be)$, if its density function is
    $$
    p(x\mid\al,\be) = \frac{\be^\al}{\Gamma(\al)}\,x^{-(\al+1)}\,\exp{\{-\be/x \}},\quad x>0\,.
    $$
    
    \item Beta:
    
    A random variable $X$ has a Beta distribution with parameters $\al,\be > 0$, denoted by $X\mid\al,\be\sim\textsf{Beta}(\al,\be)$, if its density function is
    $$
    p(x\mid\al,\be) = \frac{\Gamma(\al+\be)}{\Gamma(\al)\,\Gamma(\be)}\,x^{\al-1}\,(1-x)^{\be-1},\quad 0<x<1\,.
    $$

    \item Dirichlet: 
    
    A $K\times 1$ random vector $\Xv = (X_1,\ldots, X_K)$ has a dirichlet distribution with parameter vector $\bs{\alpha} = (\al_1,\ldots , \al_K)$, where each $\al_k > 0$, denoted by $\Xv\mid\bs{\alpha}\sim\textsf{Dir}(\bs{\alpha})$, if its density function is
    $$
    p(x\mid\bs{\alpha}) =
    \left\{
      \begin{array}{ll}
        \frac{\Gamma\left(\sum_{k=1}^K\al_k\right)}{\prod_{k=1}^K\Gamma(\al_k)}\prod_{k=1}^K x_k^{\al_k-1}, & \hbox{if $\sum_{k=1}^K x_k = 1$;} \\
        0, & \hbox{otherwise.}
      \end{array}
    \right.
    $$
    
    \item Categorical:

    A $K\times 1$ random vector $\Xv = (X_1,\ldots, X_K)$ has a categorical distribution with parameter vector $\bs{p} = (p_1,\ldots , p_K)$, where $\sum_{k=1}^K p_k =1$, denoted by $\Xv\mid\bs{p}\sim\textsf{Cat}(\bs{p})$, if its probability mass function is
    $$
    p(x\mid\bs{p}) =
    \left\{
      \begin{array}{ll}
           \prod_{k=1}^K p_k^{\ind{x = k}}, & \hbox{if $\sum_{k=1}^K x_k = 1$;} \\
        0, & \hbox{otherwise.}
      \end{array}
    \right.
    $$
    
\end{itemize}

\end{document}